\documentclass[10pt]{article}

\usepackage{amsmath}
\usepackage{amssymb}

\usepackage{graphicx}

\usepackage{cite}

\usepackage{color}

\usepackage[labelfont=bf,labelsep=period,justification=raggedright]{caption}

\makeatletter
\renewcommand{\@biblabel}[1]{\quad#1.}
\makeatother

\long\def\remove#1{}
\newcommand{\commentout}[1]{}

\begin{document}
\begin{flushleft}
{\Large
\textbf{Social Contagion: An Empirical Study of Information Spread on Digg and Twitter Follower Graphs\\}
}

Kristina Lerman$^{1,2,\ast}$,
Rumi Ghosh$^{2}$,
Tawan Surachawala$^{2}$
\\
\bf{1} USC Information Sciences Institute, Marina Del Rey, CA, USA
\\
\bf{2} Computer Science Dept., University of Southern California, Los Angeles, CA, USA
\\
$\ast$ E-mail: lerman@isi.edu
\end{flushleft}

\section*{Abstract}
Social networks have emerged as a critical factor in information dissemination, search, marketing, expertise and influence discovery, and potentially an important tool for mobilizing people. Social media has made social networks ubiquitous, and also given researchers access to massive quantities of data for empirical analysis.  These data sets offer a rich source of evidence for studying dynamics of individual and group behavior, the structure of networks and global patterns of the flow of information on them. However, in most previous studies, the structure of the underlying networks was not directly visible but had to be inferred from the flow of information from one individual to another. As a result, we do not yet understand dynamics of information spread on networks or how the structure of the network affects it. We address this gap by analyzing data from two popular social news sites.  Specifically, we extract follower graphs of active Digg and Twitter users and track how interest in news stories cascades through the graph. We compare and contrast properties of information cascades on both sites and elucidate what they tell us about dynamics of information flow on networks.

\section*{Introduction}
Social scientists have long recognized the importance of social networks in the spread of information~\cite{Granovetter}, products~\cite{Brown87,Watts07}, and innovation~\cite{Rogers03}.
Modern communications technologies, notably email and more recently social media, have only enhanced the role of networks in marketing~\cite{Domingos01,Kempe03}, information dissemination~\cite{Wu03,Gruhl04}, search~\cite{Adamic05search}, disaster communication~\cite{mashable}, and social and political movements~\cite{Lotan11}.
%Social media has proved to be a valuable source of real-time information during crises, as witnessed by its massive use in the aftermath of the triple disaster in Japan in 2011~\cite{mashable}.
%DARPA Network Challenge\footnote{https://networkchallenge.darpa.mil} successfully tested the ability of online social networks to mobilize massive ad-hoc teams to solve real-world problems, which could potentially improve disaster response and coordination of relief efforts.
In addition to making social networks ubiquitous, social media has given researchers access to massive quantities of data for empirical analysis.  These data sets offer a rich source of evidence for studying the structure of social networks~\cite{Leskovec08www} and the dynamics of individual~\cite{Barabasi06b} and group behavior~\cite{Hogg09icwsm}, efficacy of viral product recommendation~\cite{Leskovec06}, global properties of information cascades~\cite{Liben-Nowell08pnas}, and identification of influentials~\cite{Leskovec07kdd,Ghosh10snakdd,Bakshy11}. In most of these studies, however, the structure of the underlying network was not visible but had to be inferred from the flow of information  from one individual to another. This posed a serious challenge to our efforts to understand how the structure of the network affects social dynamics and information spread.

Social media sites Digg and Twitter offer a unique opportunity to study social dynamics  on networks. Both sites have become important sources of timely information for people. The social news aggregator Digg allows users to \emph{submit} links to news stories and \emph{vote} on stories submitted by other users. On Twitter users \emph{tweet} short text messages, that often contain links to news stories or \emph{retweet} messages of others. Both sites allow users to link to others whose activity (i.e., votes and tweets) they want to follow. Both sites provide programmatic access both to data about user activity and social networks. This rich, dynamic data allows us to ask new questions about information spread on networks. How far and how fast does information spread? How deeply and how widely does it penetrate? How do people respond to new information? How does network structure affect information spread? Do some network topologies accelerate or inhibit information spread?

We address some of these questions through a large scale empirical study of the spread of information on Digg and Twitter follower graphs.
For our study we collected activity data from these websites. The Digg data set contains all popular stories submitted to Digg over a period of a month, and who voted for these stories and when. Twitter data set contains tweets with embedded URLs posted over a period of three weeks. We use URLs as markers for how information diffuses through Twitter. In addition, we extracted the follower graphs of active users on these sites. These data sets allow us to empirically characterize individual and collective dynamics and trace the flow of information on the network. We measure global properties of information flow on the two sites and compare them to each other. In addition to using standard measure such as size, depth and breadth of spread, we define a new metric that characterizes how closely knit the network is through which information is spreading. We find that while characteristics of information flow on Digg and Twitter are for the most part similar, they are dramatically different from an earlier study of the structure of large scale information spread.

\section*{Description of Data}
\label{sec:social_news}
%Social media has become an important channel for people to share information. On Digg, Twitter, Slashdot, Reddit, and Facebook, among many others, users post news or links to news stories, discuss them, and share their opinions in real time. Often, these sites are the first to break important news.
%Many social media sites aggregate activity of many users to select what they deem to be the most important information feature on its front page (Digg) or as trending topics (Twitter). In addition to showcasing popular content, social media sites often enable users to discover information through their social networks, by seeing what stories their friends discovered recently. Social networks, therefore, play a critical role in information spread on these sites.

Social news aggregator {Digg} (http://digg.com) is one of the first successful social media sites. At the height of its popularity in 2007--2009, it had over 3 million registered users.  Digg allows users to submit links to and rate news stories by voting on, or \emph{digging}, them. At the time data was collected, there were many new submissions every minute, over 16,000 a day.
A newly submitted story went to the \emph{upcoming} stories list, where it remained for 24 hours, or until it was promoted to the \emph{front page} by Digg, whichever came first. Newly submitted stories were displayed as a chronologically ordered list, with the most recent story at the top of the list, 15 stories to a page.
Promoted (or `popular') stories were also displayed in a reverse chronological order on the front pages, 15 stories to a page, with the most recently promoted story at the top of the list.

Digg picked about a hundred stories daily to feature on its front page. Although the exact promotion mechanism was kept secret, it appeared to take into account the number and the rate at which story receives votes. Digg's success was largely fueled by the emergent front page, created by the collective decisions of its many users.
The importance of being promoted has, among other things, spawned a black market\footnote{As an example, see http://subvertandprofit.com} which claims the ability to manipulate the voting process.

Digg also allowed users to designate other users as friends and track their activities. The \emph{friends interface} allows users to see the stories friends recently submitted or voted for. The friendship relationship is asymmetric. When user $A$ lists user $B$ as a \emph{friend}, $A$ can watch the activities of $B$ but not vice versa. We call $A$ the \emph{fan} or a \emph{follower} of $B$. A newly submitted story is visible in the upcoming stories list, as well as to submitter's followers through the friends interface. With each vote it also becomes visible to voter's followers. The friends interface can be accessed by clicking on \emph{Friends Activity} tab at the top of any Digg page. In addition, a story submitted or voted on by user's friends receives a green ribbon on the story's Digg badge, raising its visibility to followers.

{Twitter} (http://twitter.com) is a popular social networking site that allows registered users to post and read short text messages (at most 140 characters), which may contain URLs to online content, usually shortened by a URL shortening service such as bit.ly or tinyurl. A user can also retweet the content of another user's post, sometimes prepending it with a string ``RT @\emph{x},'' where $x$ is a user's name. Like Digg, Twitter allows users to designate other users as friends and follow their tweeting activity.

\begin{figure*}[bth]
\centering{
  \begin{tabular}{cc}
  \includegraphics[height=2.3in]{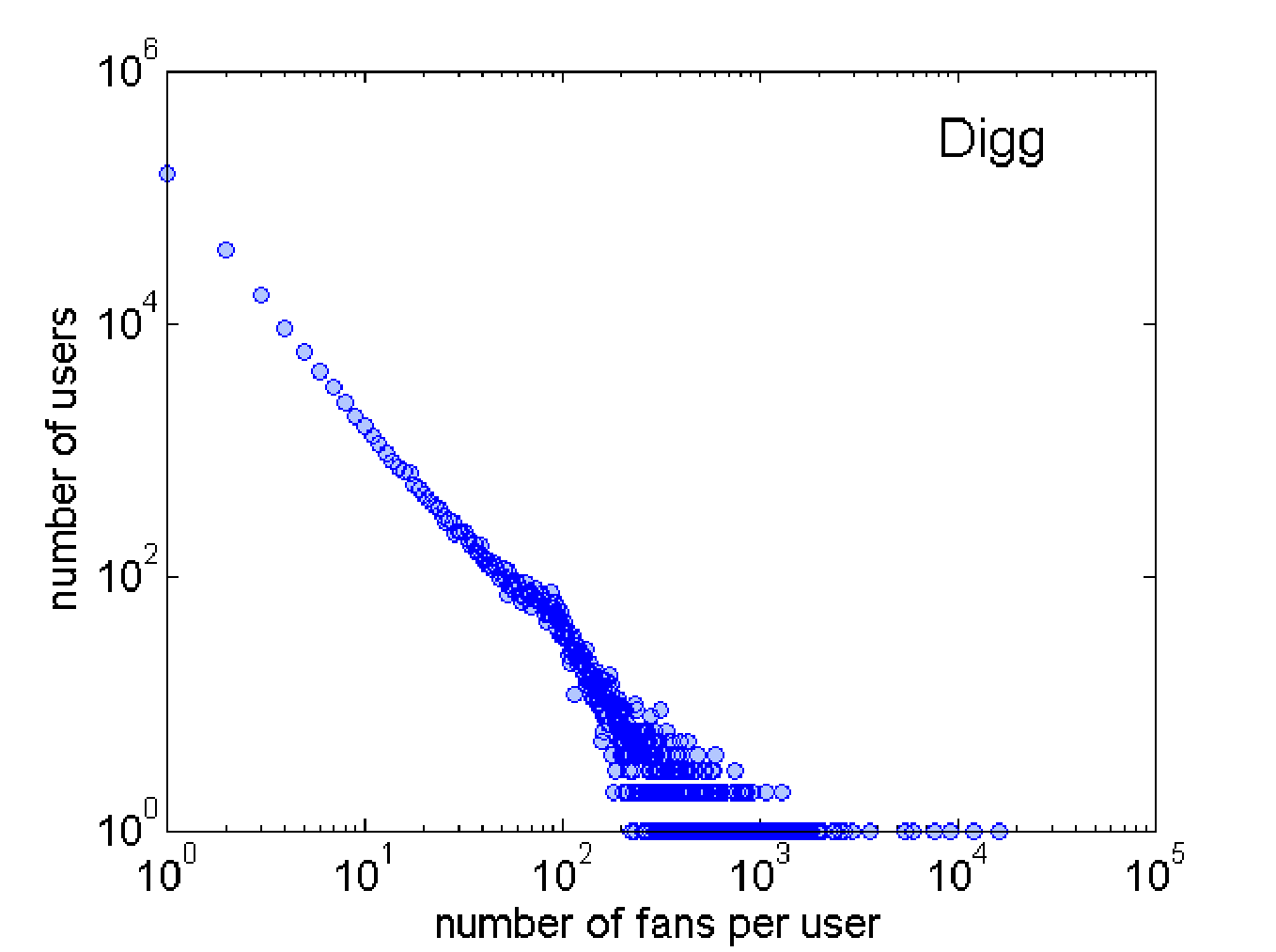}  &
  \includegraphics[height=2.3in]{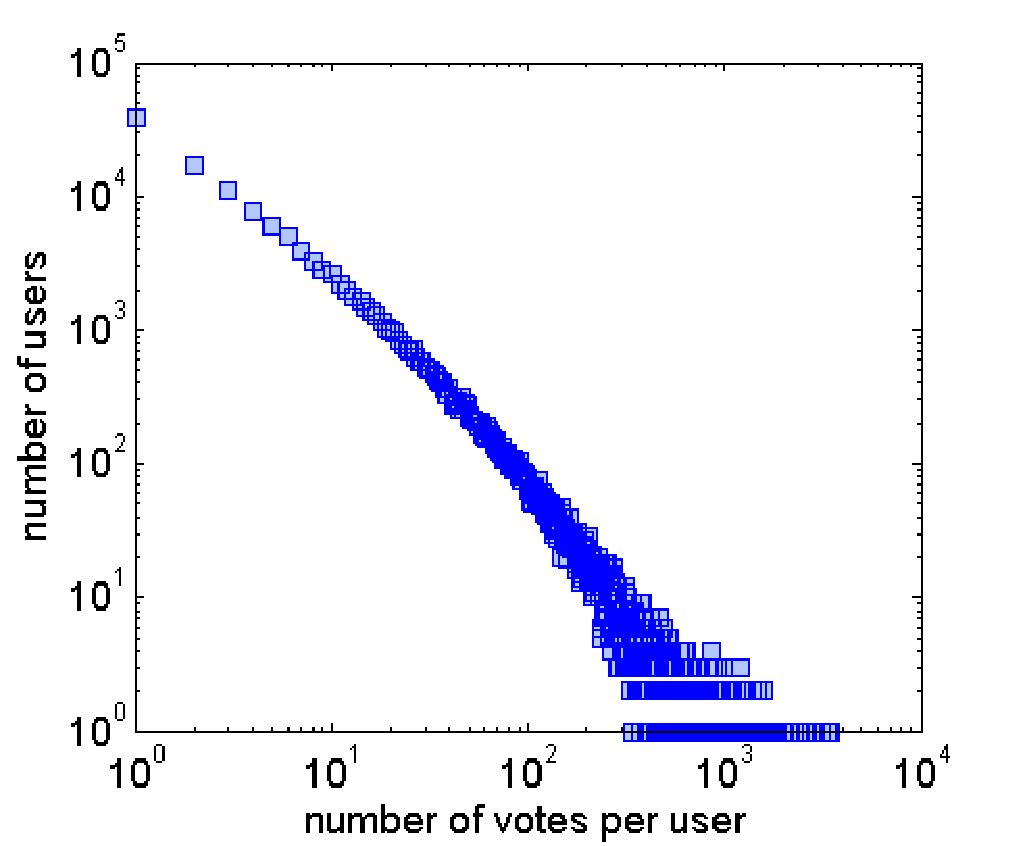}  \\
  (a) fans distribution & (b) vote distribution \\
  \includegraphics[height=2.3in]{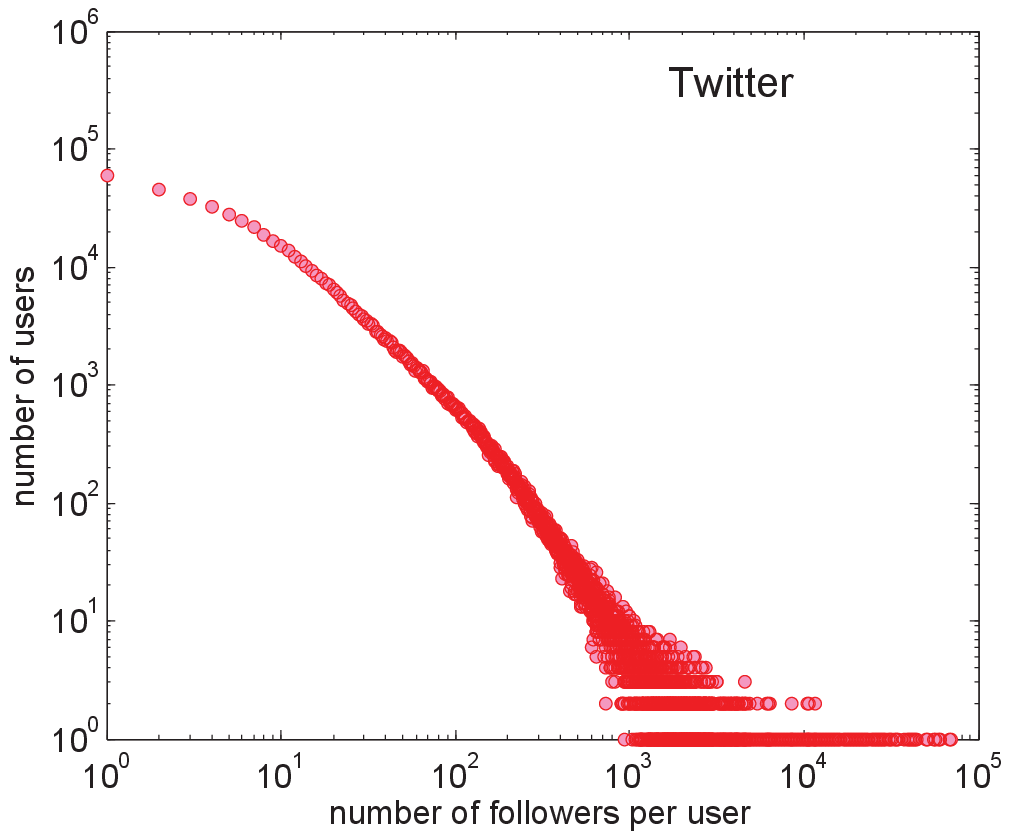}&
  \includegraphics[height=2.3in]{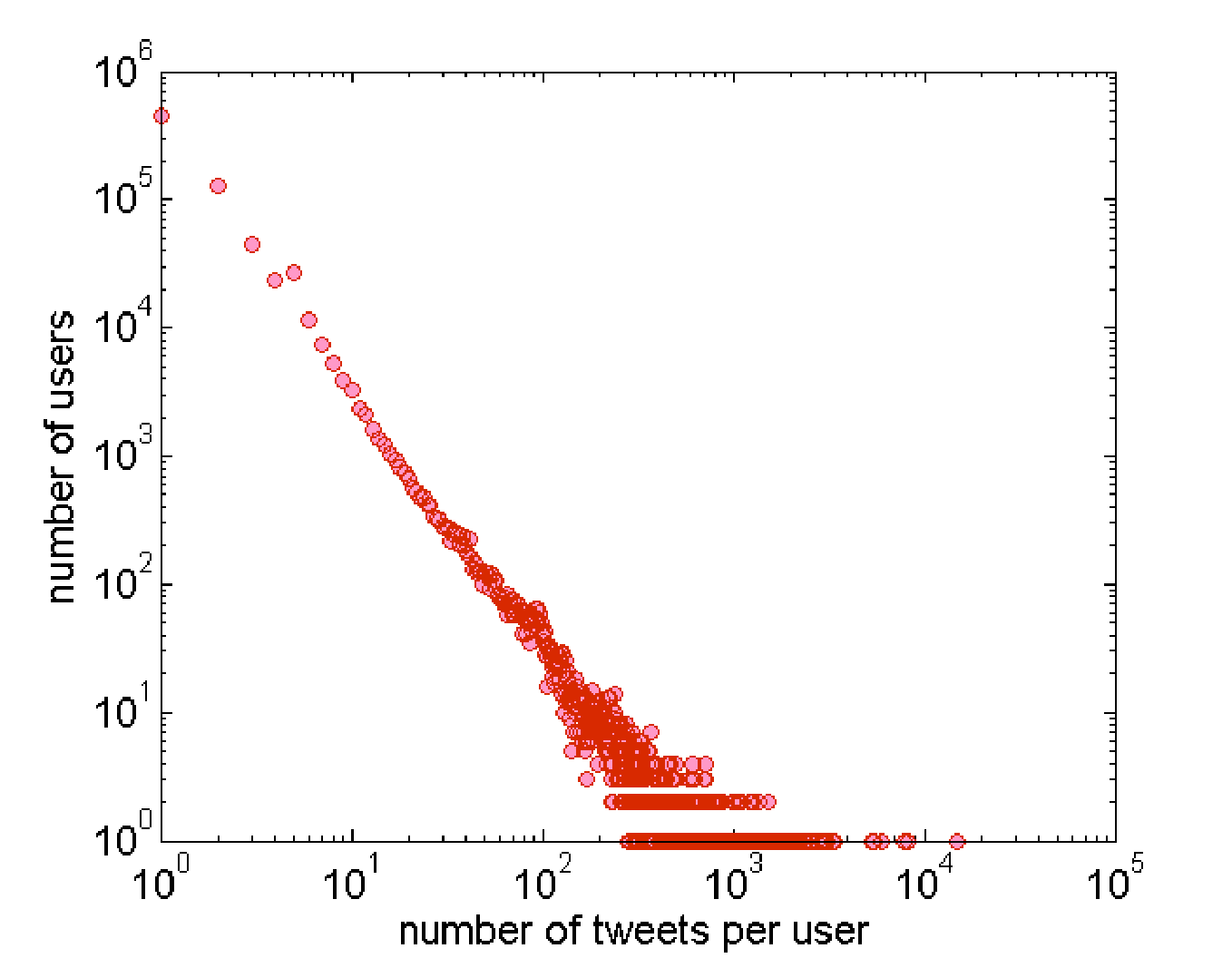}\\
 (c) followers distribution & (d) retweet distribution
  \end{tabular}
  }
  \caption{Characteristics of user activity on Digg and Twitter. Distribution of the number of (a) fans per user and (b) votes per user on Digg. Distribution of the number of (c) followers per user and (d) retweets per user on Twitter. }\label{fig:activity}
\end{figure*}

%\subsection*{Description of Data Sets}
We used Digg API to collect complete (as of July 2, 2009) voting histories of all stories promoted to the front page of Digg in June 2009.\footnote{The data set is available at http://www.isi.edu/$\sim$lerman/downloads/digg2009.html} The data associated with each story contains story id, submitter's id, list of voters with time of each vote. We also collected the time each story was promoted to the front page. In total, the data set contains over 3 million votes on 3,553 promoted stories.

%258,220 friends in friends table
% max 996 friends; min 1
% max fans: inactive(16216), kevinrose(12038), ofa(9291)
%We define an \emph{active user} as any user who voted for at least one story on Digg during the data collection period.
%We define as \emph{active user} any user who voted for at least one story on Digg.% or retweeted at least one URL on Twitter.
Of the 139,409 voters in our data set, more than half designated at least one other user as a friend. We extracted the friends of these users and reconstructed the follower graph of active users, i.e., a directed graph of active users who are following activities of other users. This graph contained 70K nodes and more than 1.7 million edges.

% data retrieval
At the time of data collection Twitter's Gardenhose streaming API provided access to a portion of real time user activity, roughly 20\%-30\% of all user activity.
We used this API to collect tweets over a period of three weeks. We focused on tweets that included a URL in the body of the message, usually shortened by some service, such as bit.ly or tinyurl. In order to ensure that we had the complete tweeting history for each URL, we used Twitter's search API to retrieve all activity for that URL. Then, for each tweet, we used the REST API to collect friend and follower information for that user.

% Twitter
%17,162,225/2 mutual links
%36,743,448 total links

%3,424,033 tweets which mentioned 70,343 distinct shortened URLs, 699,985 nodes and 36,743,448  edges
Data collection process resulted in more than 3 million tweets which mentioned 70,343 distinct shortened URLs.  There were 815,614 users in our data sample, but we were only able to retrieve follower information for some of them, resulting in a graph with almost 700K nodes and over 36 million edges.

%\subsection{Data Statistics}
Figure~\ref{fig:activity}(a) shows the distribution of number of active followers per user on Digg, while Fig.~\ref{fig:activity}(b) shows the distribution of activity, i.e., number of votes per user.
Figure~\ref{fig:activity}(c)--(d) shows the distribution of the number of followers and the number of retweets per user on Twitter.
%While the distribution of followers deviates from a true long-tailed distribution, distribution of retweeting activity (Fig.~\ref{fig:twitter}(b)) looks similar to Digg, although with a higher exponent.
%The difference in slopes in these distribution is likely explained by the level of effort~\cite{Wilkinson08} required to vote on Digg vs retweet on Twitter.
The ``heavy tailed'' distribution of voting and retweeting are typical of social production and consumption of content. In a heavy-tailed distribution a small but non-vanishing number of items generate uncharacteristically large amount of activity. While the overwhelming majority of Digg users cast fewer than 10 votes, a handful of users voted on thousands of stories over the period of a month, or hundreds of stories a day. Similarly, on Twitter a handful of users retweeted thousands of URLs. In addition to Digg~\cite{Wu07} and Twitter, long tailed distributions have been observed in voting on Essembly~\cite{HoggSzabo09icwsm}, edits of Wikipedia articles~\cite{Wilkinson08}, and music downloads~\cite{Salganik06} and other and real-world complex networks~\cite{Clauset09}.
%Log-normal distributions are a result of a multiplicative process~\cite{Mitzenmacher04}.
Understanding the origin of such distributions is the next challenge in modeling user activity on social media sites.

\section*{Results}
%\subsection*{Dynamics of Information Spread}
\label{sec:dynamics_voting}

Our data sets contain the record of all votes on Digg's front page stories and retweets of URLs on Twitter, from which we can  reconstruct dynamics of information spread. In addition to voting history, we also know the active follower graph of Digg and Twitter users, and use it to study how interest in stories spreads through the social networks of Digg and Twitter.

\begin{figure*}[tbhp]
\centering{
  \begin{tabular}{cc}
  \includegraphics[width=2.5in]{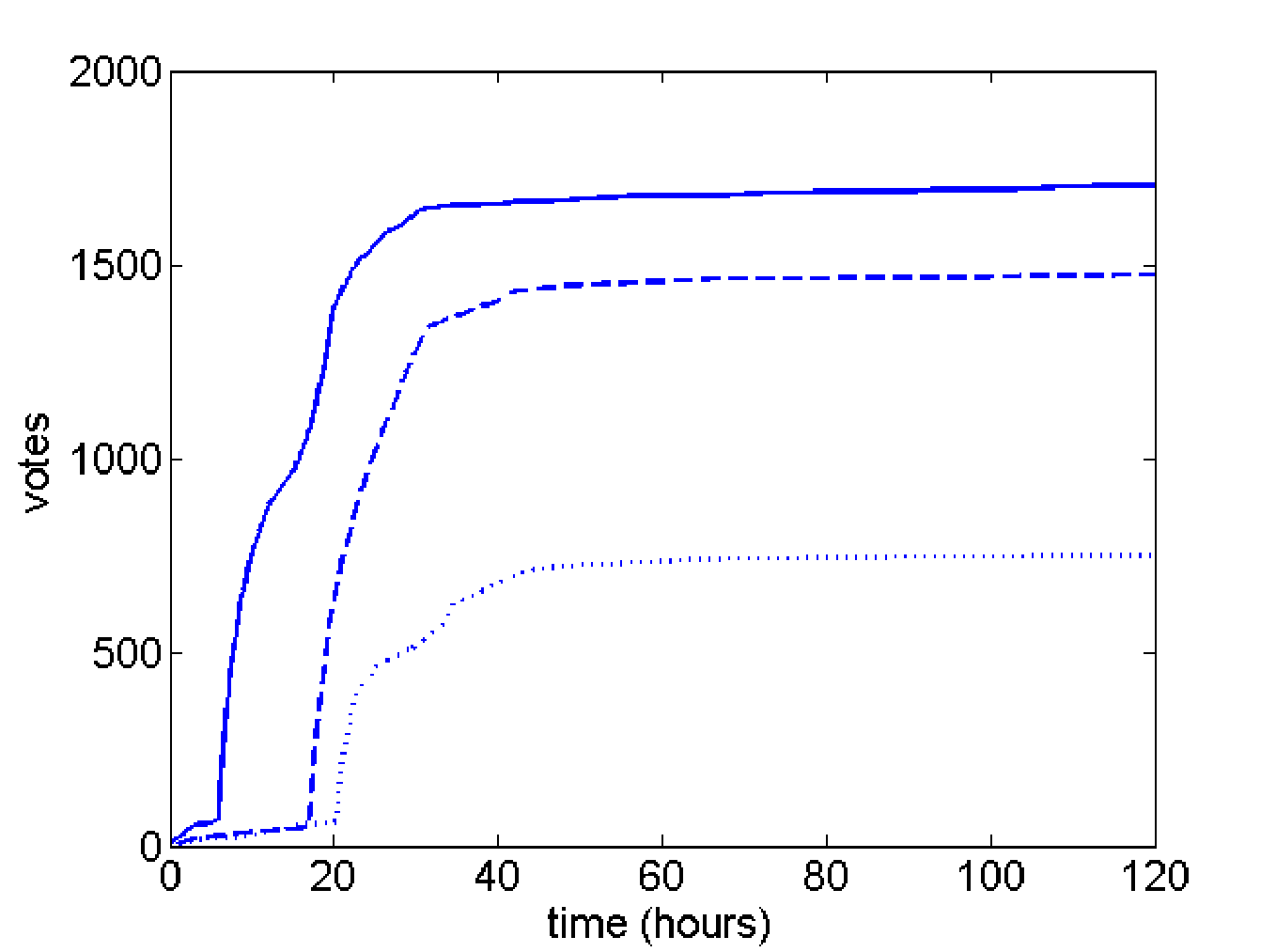} &
    \includegraphics[width=2.5in]{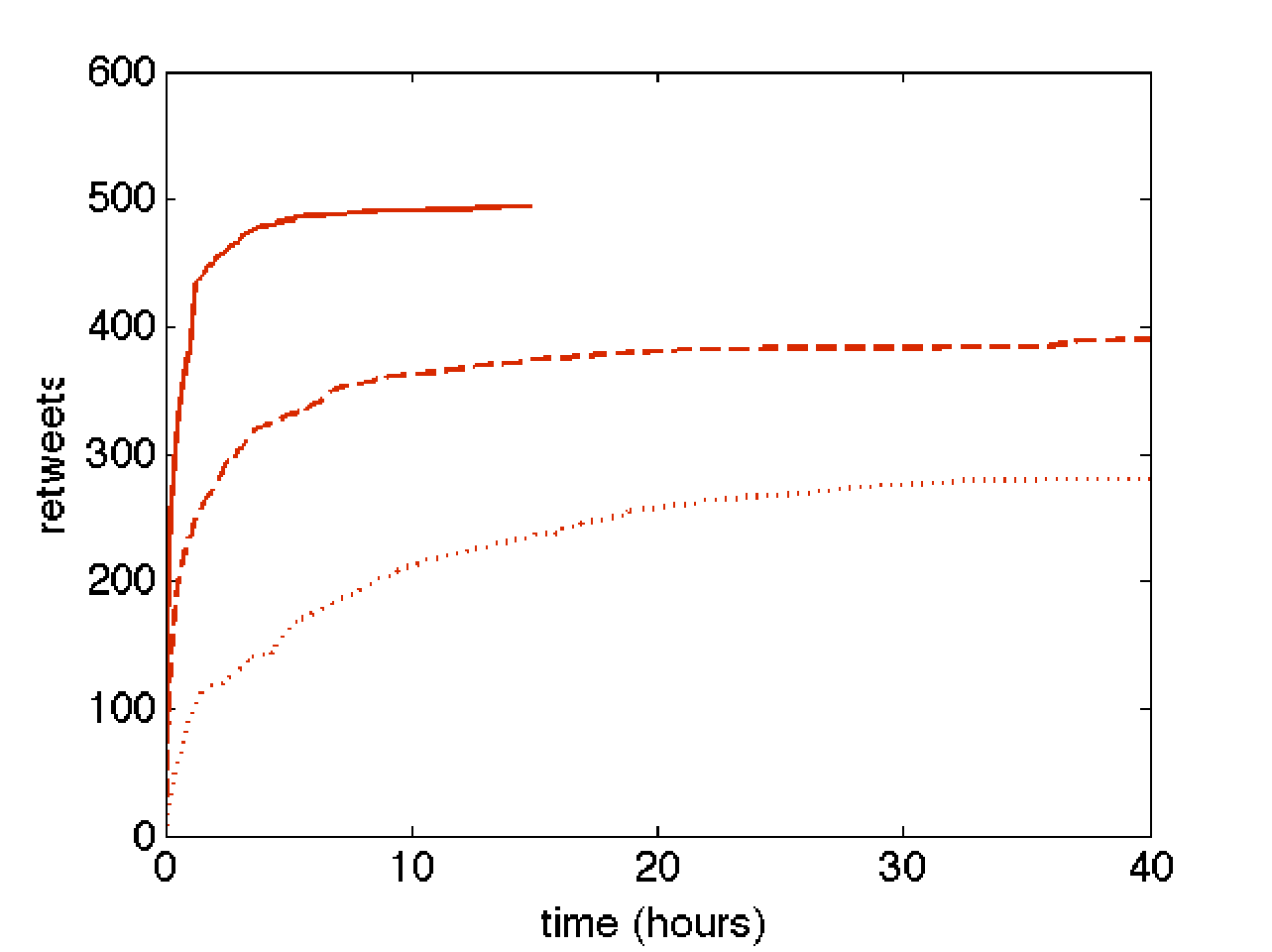}\\
     (a) Digg &    (b) Twitter
  \end{tabular}
}
  \caption{Dynamics of popularity on Digg and Twitter. (a) Number of votes received by three stories on Digg since submission. (b) Number of times stories were retweeted since the first post vs time.
}\label{fig:dynamics}
\end{figure*}

\subsection*{Evolution of Popularity}
Figure~\ref{fig:dynamics} shows the evolution of the number of votes received by three stories on Digg and the number of times URLs to three news stories were retweeted on Twitter.
Although the details of the dynamics differ from story to story, the general features of the evolution of popularity are shared by all stories. The evolution of story on Digg, Figure~\ref{fig:dynamics} (a), has two distinct phases: the \emph{upcoming} phase and the \emph{promoted} phase.  While in the upcoming stories queue, a newly submitted story accumulates votes at some slow rate as seen in the initial upcoming phase. The point where the slope abruptly changes corresponds to promotion to the front page and the beginning of the promoted phase. After promotion the story is visible to many more people who only visit Digg's front page, and the number of votes grows at a much faster rate. As the story ages, accumulation of new votes slows down and saturates.
These dynamics are well characterized by a model of user behavior~\cite{Hogg09icwsm,Lerman10www} that takes into account visibility of stories through the Digg user interface and how interesting they are to users.

%The total number of votes the story receives by that time gives a measure of its success or {popularity}.
\commentout
{
There is no differentiation on Twitter between ``upcoming'' and ``promoted'' stories. Therefore, popularity of news stories and blog posts on Twitter grows smoothly until saturation~\cite{Ghosh11snakdd}.
On both sites, it takes a day, or less, for the number of votes/retweets to saturate to their final values. After a day or two, it is unlikely a story will get new votes.}

In contrast to Digg, the evolution of story popularity on Twitter cannot be broken down into two distinct phases. This is probably because content spreads primarily through the follow graph and no mechanism of promotion exists on Twitter. Therefore, popularity of news stories and blog posts on Twitter grows smoothly until saturation~\cite{Ghosh11snakdd}.

 On both sites, it takes a day, or less, for the number of votes/retweets to saturate to their final values. After a day or two, it is unlikely a story will get new votes.

\begin{figure*}[tbh]
\begin{center}
 \begin{tabular}{cc}
  \includegraphics[width=2.5in]{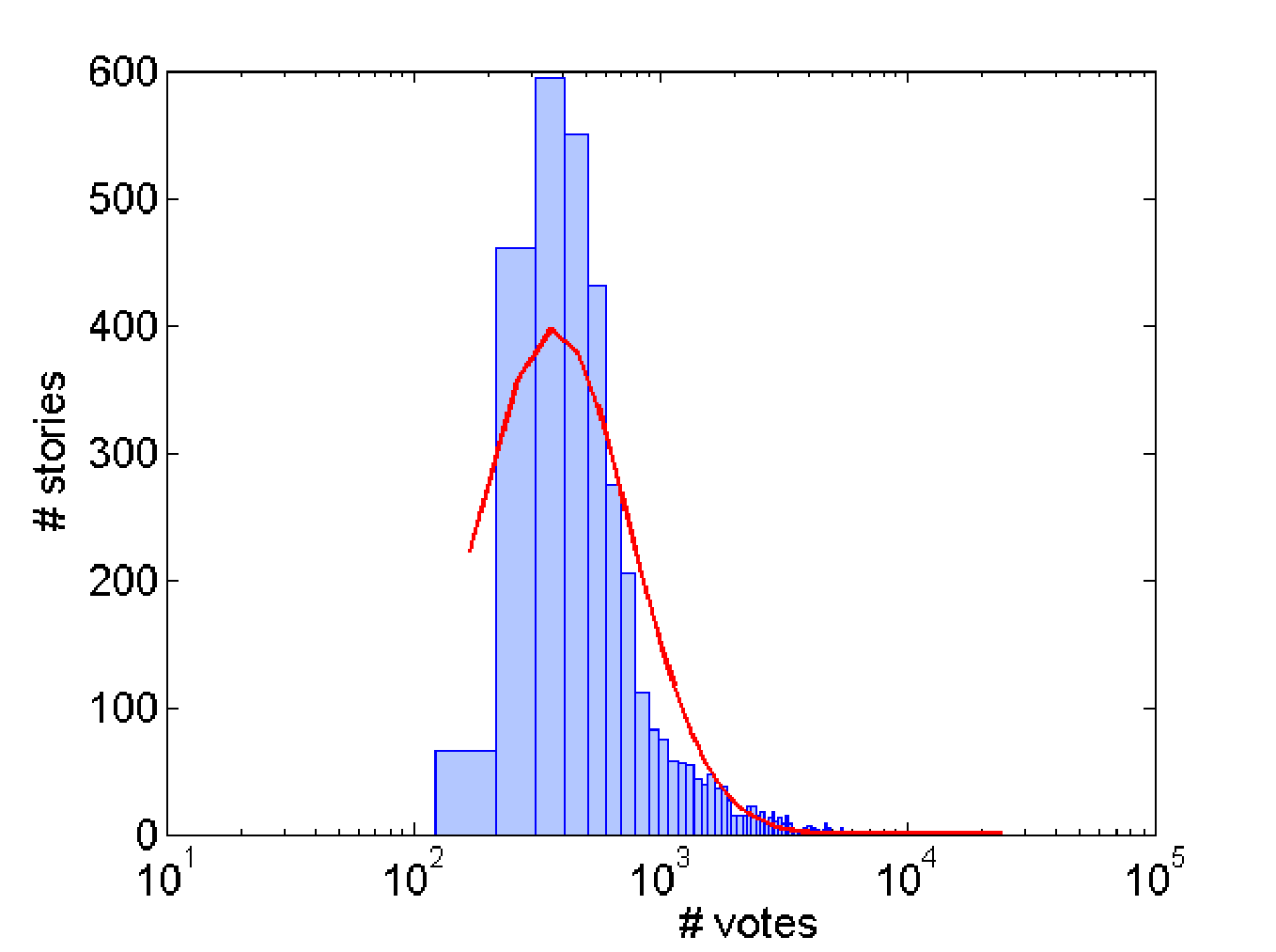} &
  \includegraphics[width=2.5in]{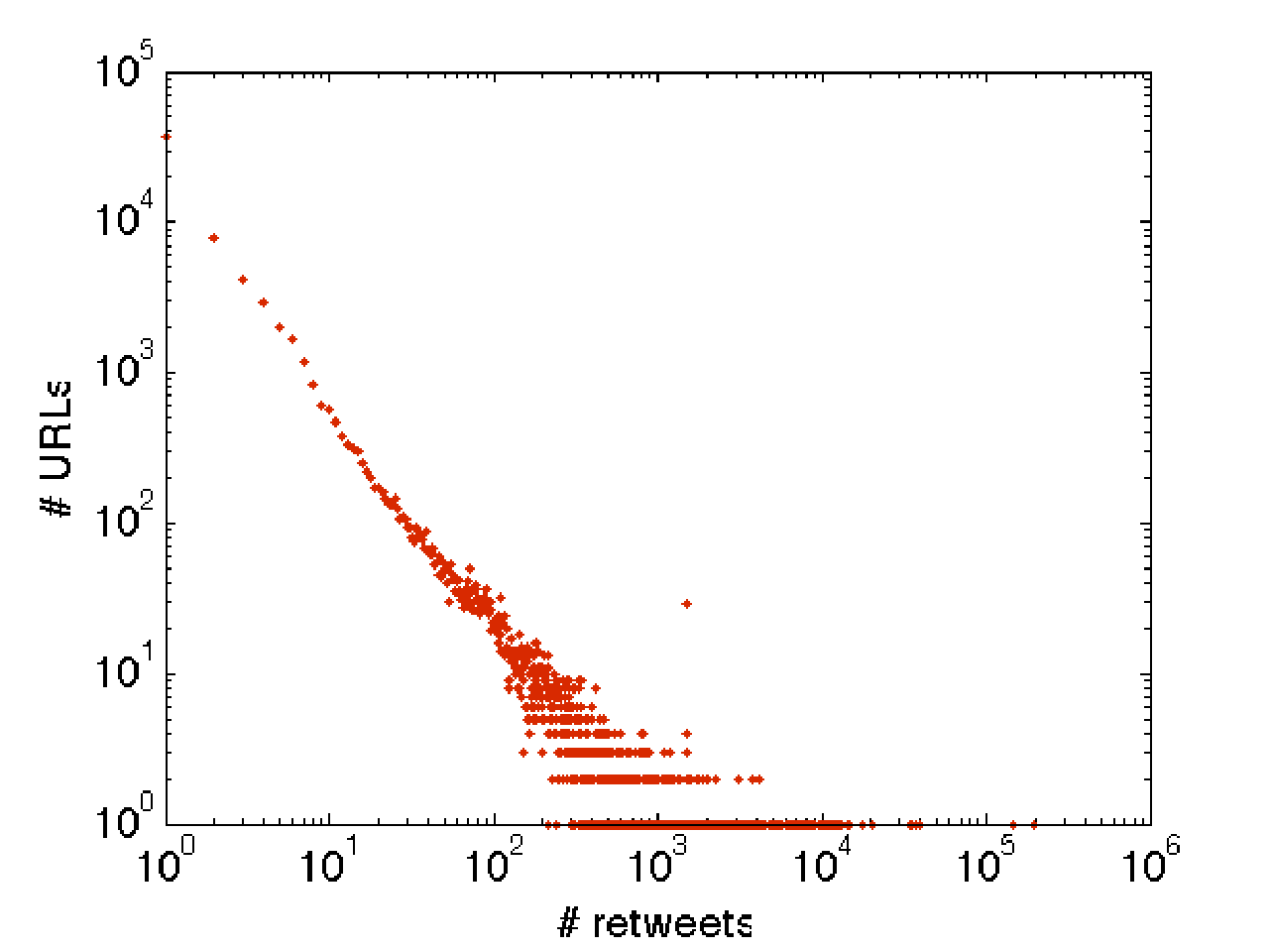} \\
  (a) Digg &   (b) Twitter
  \end{tabular}
\end{center}
  \caption{Distribution of content popularity. (a) Distribution of the total number of votes received by Digg stories, with line showing log-normal fit. (b) Distribution of the total number of times stories in the Twitter data set were retweeted. }\label{fig:votes_histo}
\end{figure*}

\remove{
\begin{figure*}[tbh]
\begin{center}
  \begin{tabular}{cc}
  \includegraphics[width=2.5in]{figs/fanvotes_distribution_with_fit}
  &
  \includegraphics[width=2.5in]{figs/fantweets_histo} \\
  (a) Digg &
  (b) Twitter
  \end{tabular}
\end{center}
  \caption{Distribution of story cascade sizes. (a) Histogram of the distribution of the total number of fan votes received by Digg stories (size of the interest cascade). The inset shows the distribution of the number of votes from submitter's fans. (b) Histogram of the distribution of the total number of retweets from followers. The inset shows the distribution of the number of retweets of a story from submitter's followers. }\label{fig:fanvotes_histo}
\end{figure*}
}

%\subsection*{Distribution of popularity}
The total number of times the story was voted for or retweeted reflects its popularity among Digg and Twitter users respectively.
The distribution of story popularity on either site, Figure~\ref{fig:votes_histo}, shows the `inequality of popularity'~\cite{Salganik06}, with relatively few stories becoming very popular, accruing thousands of votes (retweets), while most are much less popular, receiving a few hundred votes (retweets).

\begin{figure}[tbh]
\begin{center}
\includegraphics[width=3in]{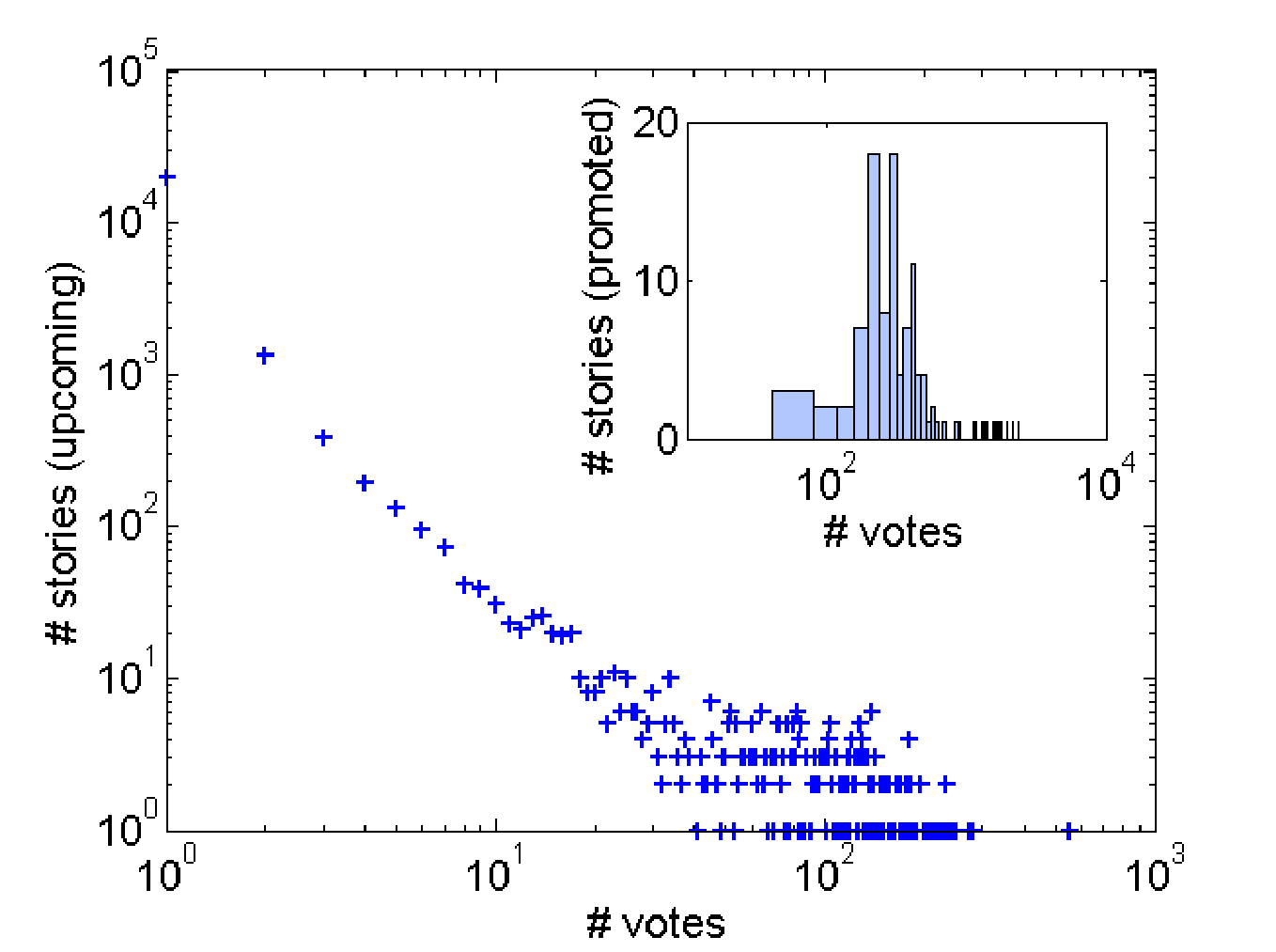}
\end{center}
  \caption{Distribution of the number of votes received by upcoming stories on Digg. Inset shows the distribution of the number of final votes received by a few of the upcoming stories that were eventually promoted by Digg.}\label{fig:upcoming}
\end{figure}

There is a striking difference between distributions of story popularity on Digg and Twitter. The distribution of popularity on Digg is well described by a lognormal distribution (shown as the red line), with the mean of 614 votes.
There is no preferred number of retweets for URLs on Twitter, with popularity showing a power law-like behavior.

{What gives rise to the difference of popularity distribution in Digg and Twitter?}
Wu and Huberman~\cite{Wu07} proposed a phenomenological model that explained the log-normal distribution of popularity on Digg as a byproduct of competition for attention for news stories and their decaying novelty. We find that the difference is driven largely by Digg's promotion mechanism, which highlights a handful of stories on its popular front page. The test this hypothesis, in July 2010 we retrieved information about more than 20,000 stories stories submitted to Digg's upcoming stories queue over the course of one day. Figure~\ref{fig:upcoming} shows the distribution of the total number of votes received by these stories. This distribution is similar to that in Fig.~\ref{fig:votes_histo}(b). Of these stories, about 100 were promoted to the front page and their popularity continued to evolve. The inset in Fig.~\ref{fig:upcoming} shows the final popularity of the promoted stories, which resembles the log normal distribution of story popularity on Digg. Wu and Huberman model applies to front page stories, but it does not explain the power-law distribution of popularity that is observed in the absence of filtering mechanism imposed by Digg's promotion algorithm.

\subsection*{Properties of Information Cascades}
\label{sec:network-dynamics}
%Social networks play an important role in the spread of ideas and information~\cite{Rogers03,Young03,Watts07} in society. Online social networks play an equally important role in the spread of ideas through the blogosphere and email~\cite{Gruhl04,Wu04,Liben-Nowell08pnas}. Availability of large-scale, time-resolved data about user behavior in social media allows us to ask new questions about social networks and social behavior, including how information spreads on networks? How fast and how far does it spread? How does the structure of the network affect information spread, etc.? We address some of these questions below through a quantitative empirical study of the spread of information on Digg and Twitter follower networks.

A \emph{cascade} is a sequence of activations generated by a \emph{contagion process}, in which nodes cause connected nodes to be activated with some probability~\cite{Ghosh11wsdm}. In analogy with the spread of an infectious disease on a network, an \emph{infected} (activated) node \emph{exposes} his followers to the infection. Disease cascades through the network as exposed followers become infected, thereby exposing their own followers to the disease, and so on. The \emph{seed} of a cascade is the node that initiates the cascade.

\begin{figure}[tbh]
\centering{
\begin{tabular}{@{}|c@{}|@{}c|@{}}
\hline
 \includegraphics[width=1.5in]{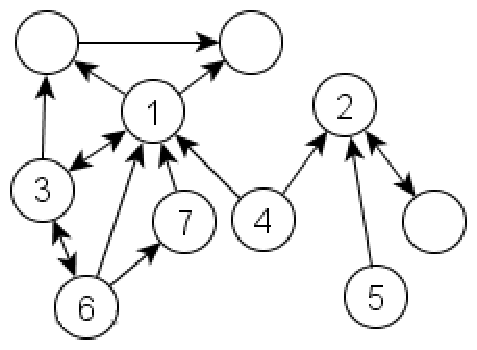} &
 \includegraphics[width=1.5in]{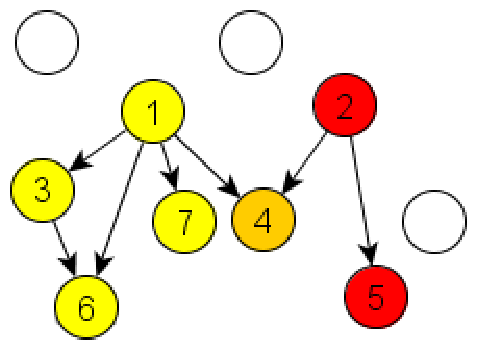}\\
 (a) & (b) \\  \hline
\end{tabular}
}
  \caption{An toy example of an information cascade on a network. Nodes are labeled in the temporal order in which they are activated by the cascade.  The nodes that are never activated are blank. (a)  The edges show the underlying follower network. Edge direction shows the semantics of the connection, i.e., nodes are watching nodes they point to. (b) Two cascades on the network (shown in yellow and red).   Node 1 is the seed of the first (yellow) cascade and node 2 is the seed of the second (red) cascade. Node 4 belongs to both cascades and is shown in orange. }
  \label{fig:network}
\end{figure}

The spread of a story through the Digg or Twitter follower graphs can be described as a contagion process on the follower graph where interest in a story spreads from voters/tweeters to their followers. We illustrate this idea with a simple example. Figure~\ref{fig:network}(a) shows a directed follower graph with link direction indicating \emph{following} relation: e.g.,  user $4$ is following activities of users $1$ and $2$. A user is infected by voting for a story. Interest in a story spreads \emph{from} infected nodes \emph{to} their followers, e.g., from users $1$ and $2$ {to} $4$. Figure~\ref{fig:network}(b) shows two cascades on the follower graph in Fig.~\ref{fig:network}(a). Users are labeled in the order they vote for a story. There are two independent seeds, namely  users $1$ and $2$.

In information cascades, the seed is an independent originator of information, who then influences others to adopt, endorse, or transmit that information.
As interest spreads, it generates multiple cascades from independent seeds.   A node can participate in more than one cascade (like user $4$ in the above example, who  participates in both cascades), resulting in a commonly observed ``collision of cascades''~\cite{Leskovec07} phenomenon.  In Digg or in Twitter, cascades may `collide' when a voter participates in more than one cascade.

We call the cascade that starts with the submitter and includes all voters who are connected either directly or indirectly to the submitter via the follower network the \emph{principal cascade} of the story. The principal cascade of the contagion process shown in Fig.~\ref{fig:network}(b) includes users $1$, $3$, $4$, $6$, and $7$.

\subsubsection*{Characterizing Information Cascades}
We can treat the evolution of each story on Digg and on Twitter as an independent contagion process, which might comprise of multiple cascades.
The following quantities are useful for quantitatively characterizing macroscopic properties of information cascades~\cite{Ghosh11wsdm}.
\begin{itemize}
\item Cascade \emph{size} is the total number of nodes infected by the seed.
\item The   \emph{maximum diameter} of the cascade is the length of the longest  chain~\cite{Leskovec07}.  The diameter of the principal cascade in Fig. \ref{fig:network}(b) is two (longest chain is $1 \rightarrow 3 \rightarrow 6$).
\item The \emph{minimum diameter} (graph diameter) of a cascade is the longest of the shortest paths from the seed to all nodes in the cascade \cite{harary}. The minimum diameter of the principal cascade in Fig.~\ref{fig:network}(b) is one.
\item The \emph{spread} of the cascade is the maximal branching number of its participants, i.e., the maximum number of users a single voter infects in a cascade. The spread of the principal cascade in Fig.~\ref{fig:network}(b) cascade is 4 and of the second (red) cascade  is 2.
\end{itemize}
\noindent For each story (contagion process) in the dataset, we measure these macroscopic properties of the principal cascades and plot their distribution over all the stories propagating in the network for both Digg and Twitter.

 In addition, to get the aggregate characteristics of all the cascades constituting a contagion process we compute the following global distributions for the contagion process:
\begin{itemize}
\item \emph{Global cascade size}: Distribution of sizes of all the cascades over all stories.
 \item  \emph{Largest cascade size}: Distribution of sizes of the  largest cascades over all stories.
\item \emph{Global maximum diameter}: Distribution of the largest of the maximum diameter of all cascades for a given story, calculated  over all stories.
\item  \emph{Global minimum diameter}: Distribution of the largest of the shortest paths of all nodes participating in the contagion process, to any seed in the contagion process or story, calculated over all stories.
\item  \emph{Global spread}: Distribution of the maximal branching of all the participants of a contagion process (story), participating in any of the cascades comprising the contagion process, calculated over all stories.
\item  \emph{Community value}: We define the \emph{community value} of the contagion process as the total number of possible activations of each node participating in the contagion process, aggregated over all the participating nodes. In other words, when information is spreading within a community, a node could have been infected by any of the infected nodes it is following. Community value measures the number of edges of activation within the contagion process and indicates how closely interconnected  are the participating nodes. The community value of the contagion process in Fig.~\ref{fig:network}(b) is seven, with  five activation edges in the yellow cascade, and two in red cascade.
\item The \emph{normalized community value} simply divides the community value by the size of the total infected or activated nodes participating in the contagion process (story). This measure gives a rough estimate of on average, how many of a voter's friends have voted on a story, before a voter  herself votes on it.
\end{itemize}
The characteristics of these observed aggregated properties of contagion process occurring on the network are indicative of how the nature of the underlying network may affect the spread of information over it.

\begin{figure*} [htbp]
\centering{
\begin{tabular}{cc}
  \includegraphics[width=2.5in]{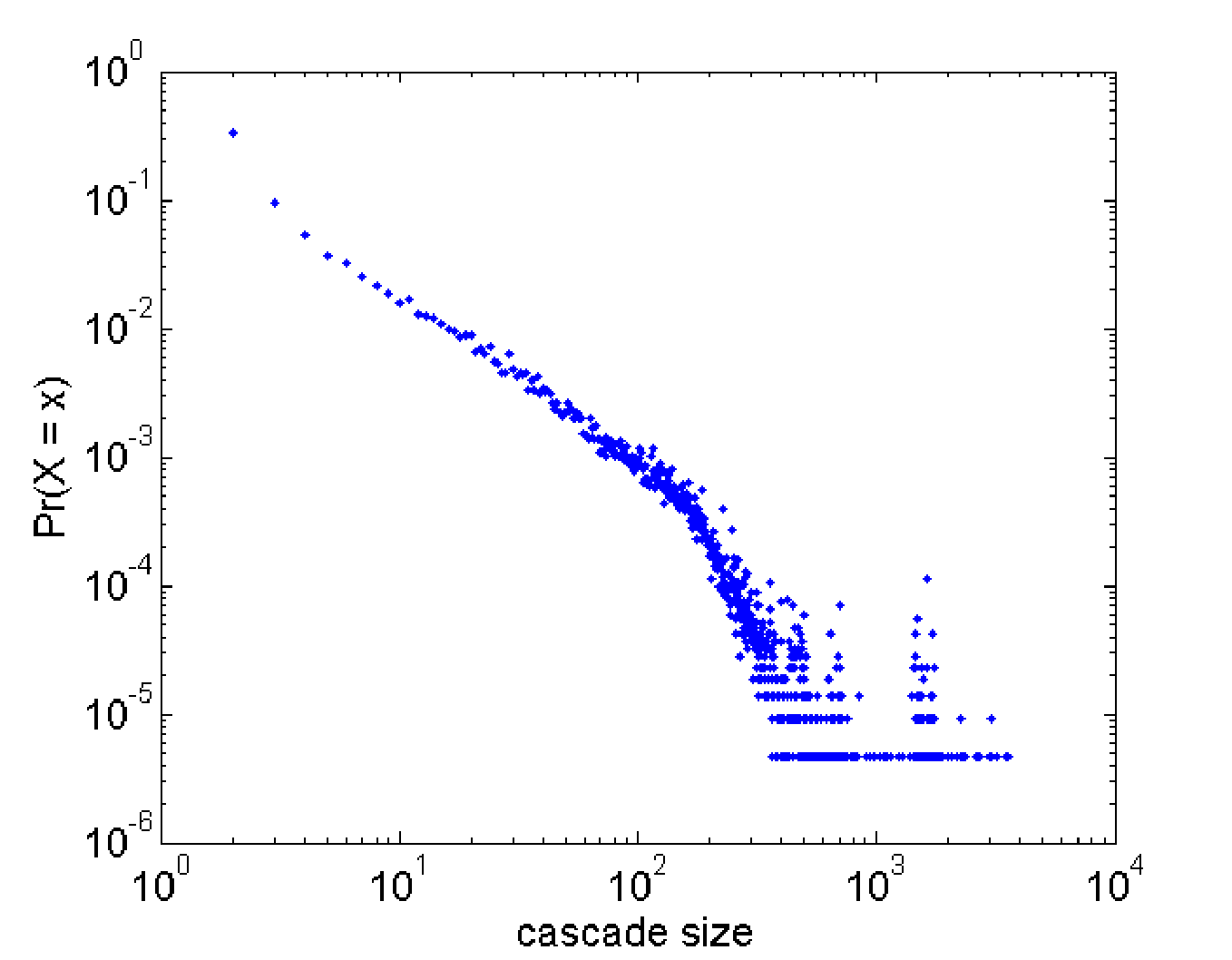} &
 \includegraphics[width=2.5in]{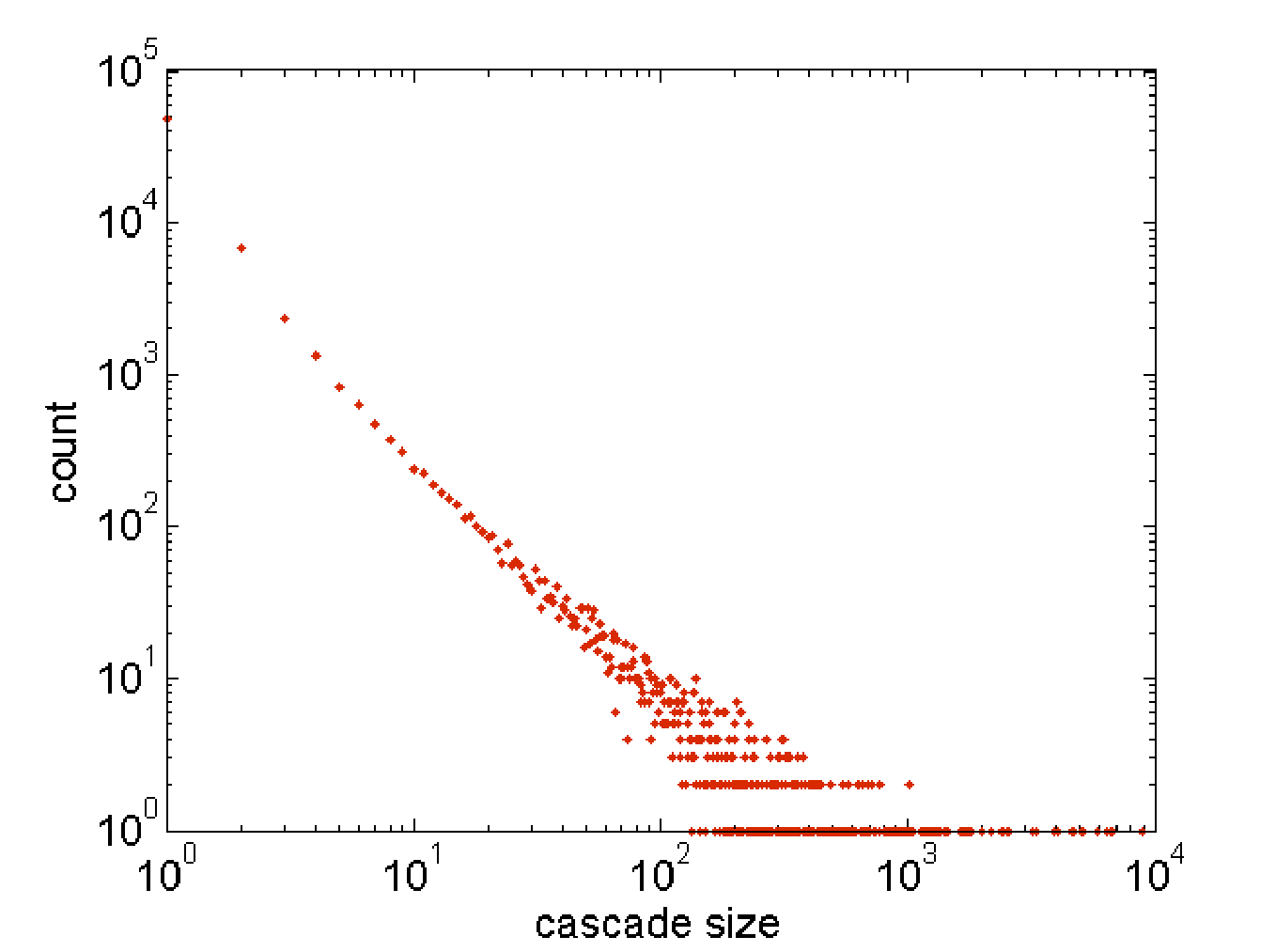} \\
     {(a) Global cascade size on Digg} &  {(b) Largest cascade size on Twitter } \\
  \includegraphics[width=2.5in]{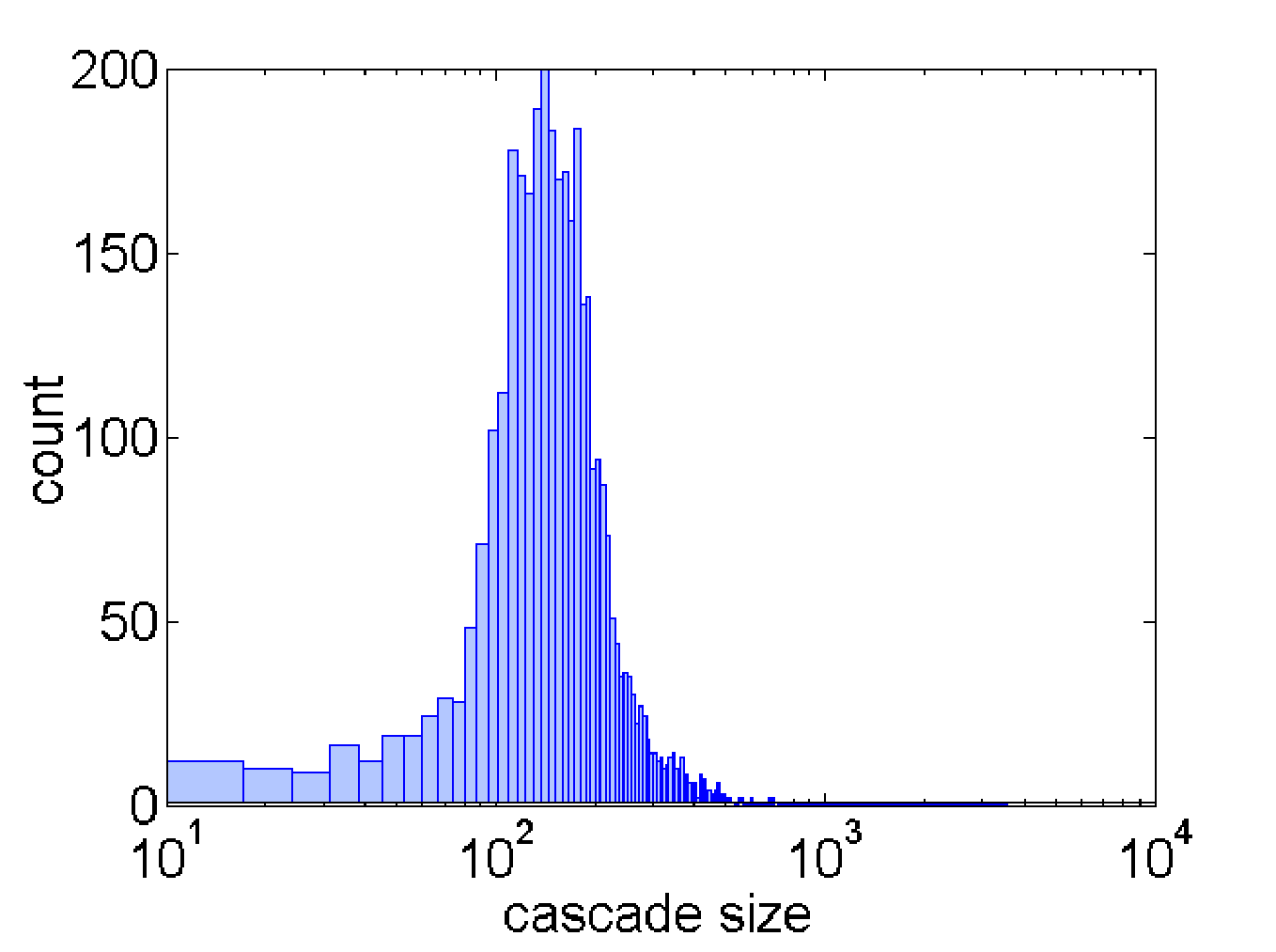} &
  \includegraphics[width=2.5in]{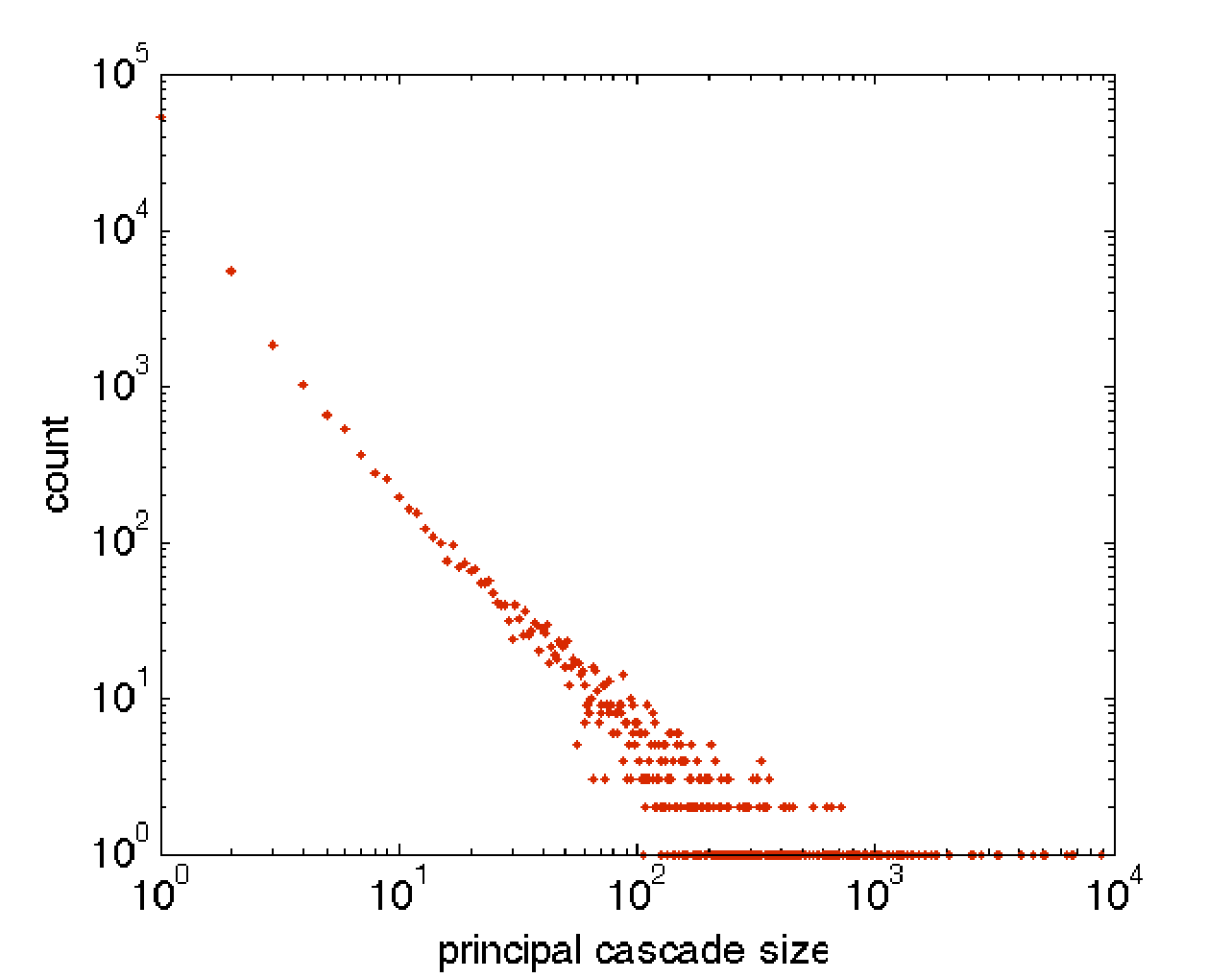} \\
     {(c) Principal cascade size on Digg } & {(d) Principal cascade size on Twitter} \\
\end{tabular}
}
\caption{Distribution of cascade sizes in Digg and Twitter }
\label{fig:cascade_size}
\end{figure*}

\subsubsection*{Cascade Size Distribution}
Given the follower graph and a time sequence of votes, we extract individual cascades generated by all Digg and Twitter stories using using the methodology described in~\cite{Ghosh11wsdm}. Figure~\ref{fig:cascade_size} shows the probability distribution of the cascade sizes. The lognormal or stretched exponential (Weibull) gives a good fit of the global cascade size for Digg, Fig.~\ref{fig:cascade_size}(a), while the power law accounts for just a small percentage at the tail of the distribution~\cite{Ghosh11wsdm}. The largest cascade size distribution on Twitter also has a similar long tail distribution.
The principal cascade size distribution on Digg takes the log normal form of the popularity distribution, with the most common size for a Digg cascade being about 200. This observation can be explained by the fact that Digg activity is dominated by top users~\cite{Lerman07sma}, who not only submit a disproportionate share of promoted stories, but are also likely to be connected to one another. We believe that the peak in the principal cascade size distribution reflects the influence of the top users.

\begin{figure*} [htbp]
\centering{
\begin{tabular}{cc}
  \includegraphics[width=2.5in]{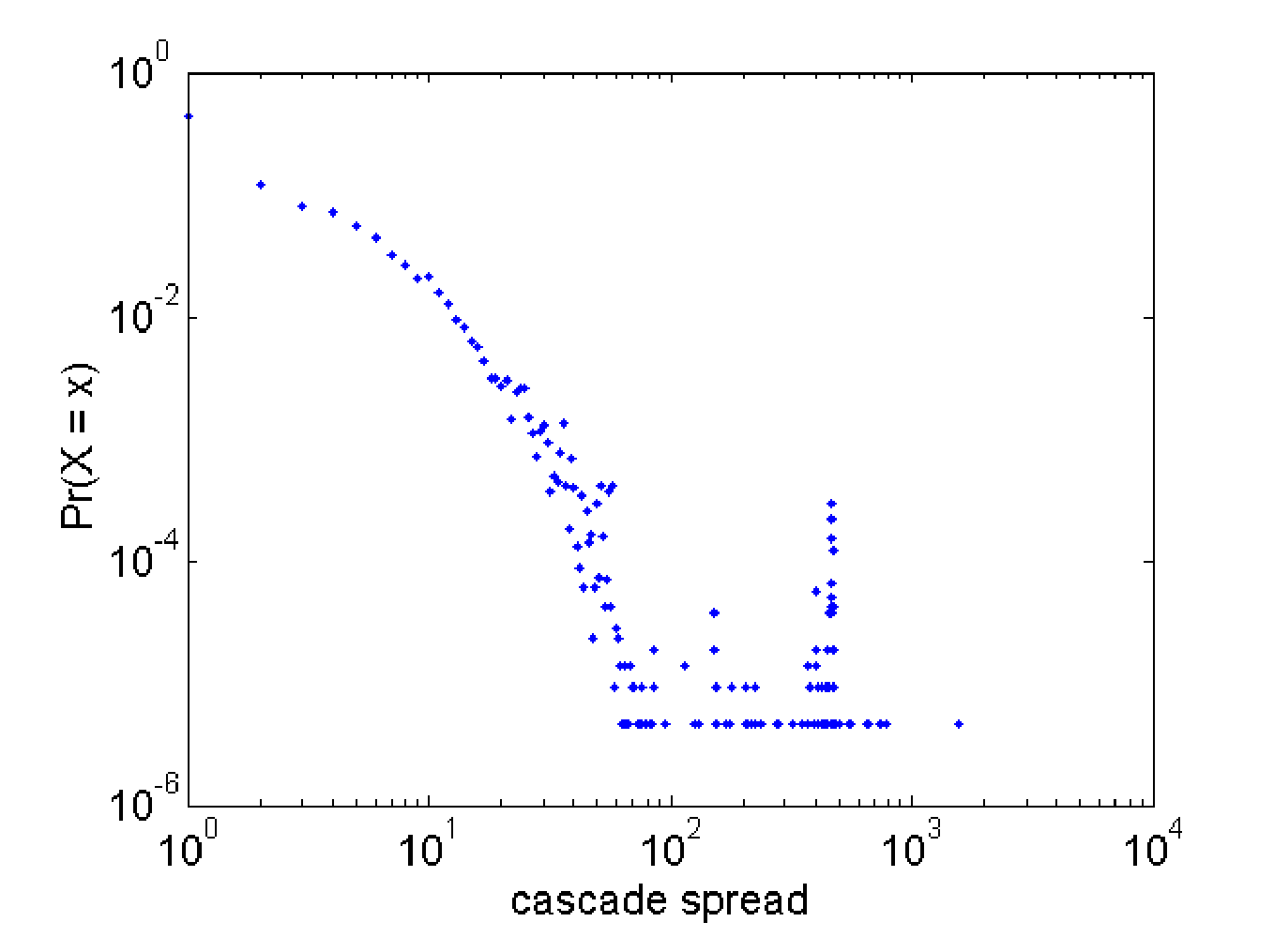} &
 \includegraphics[width=2.5in]{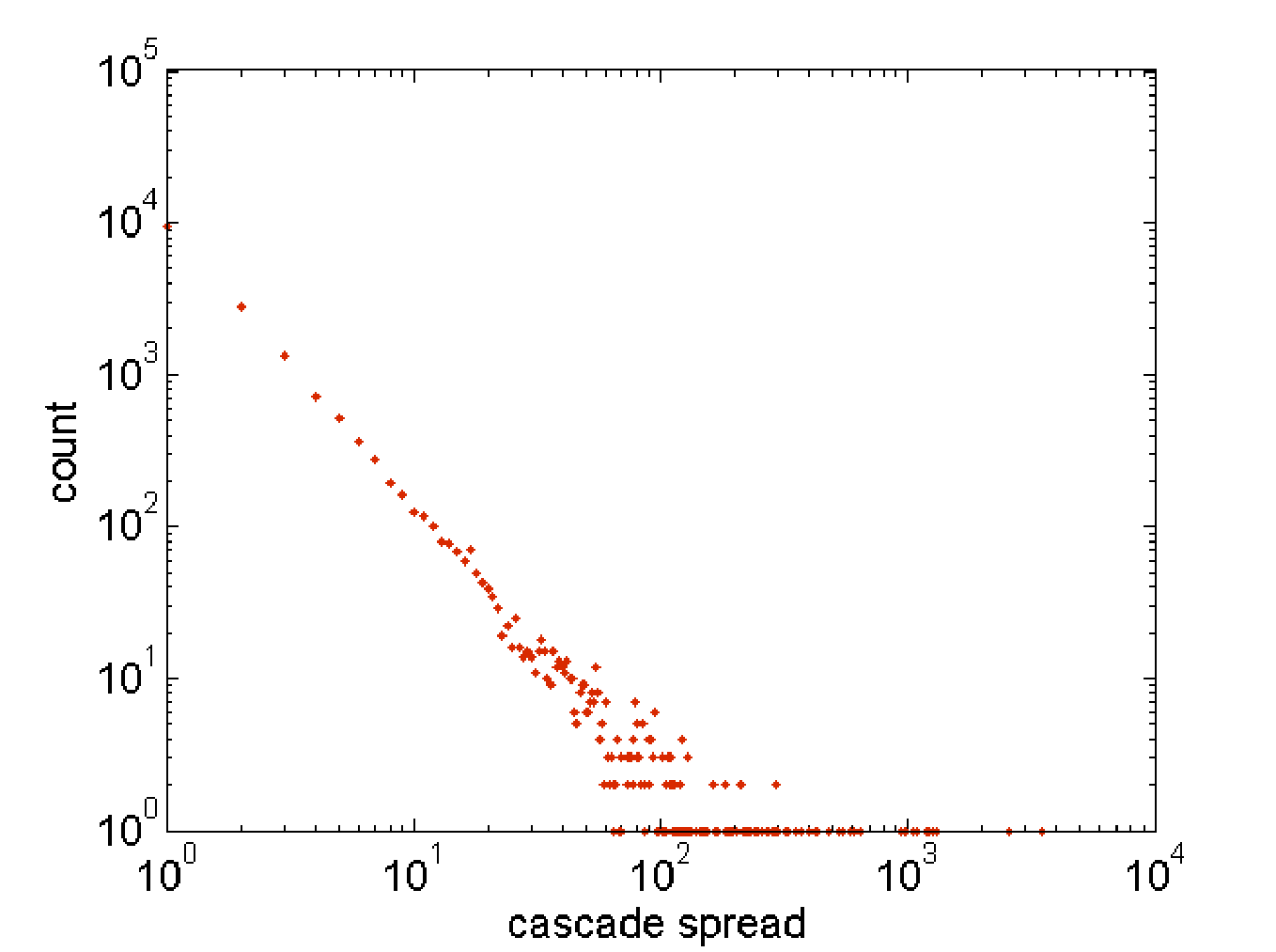} \\
     {(a) Global spread on Digg} &  {(b) Global spread on Twitter } \\
  \includegraphics[width=2.5in]{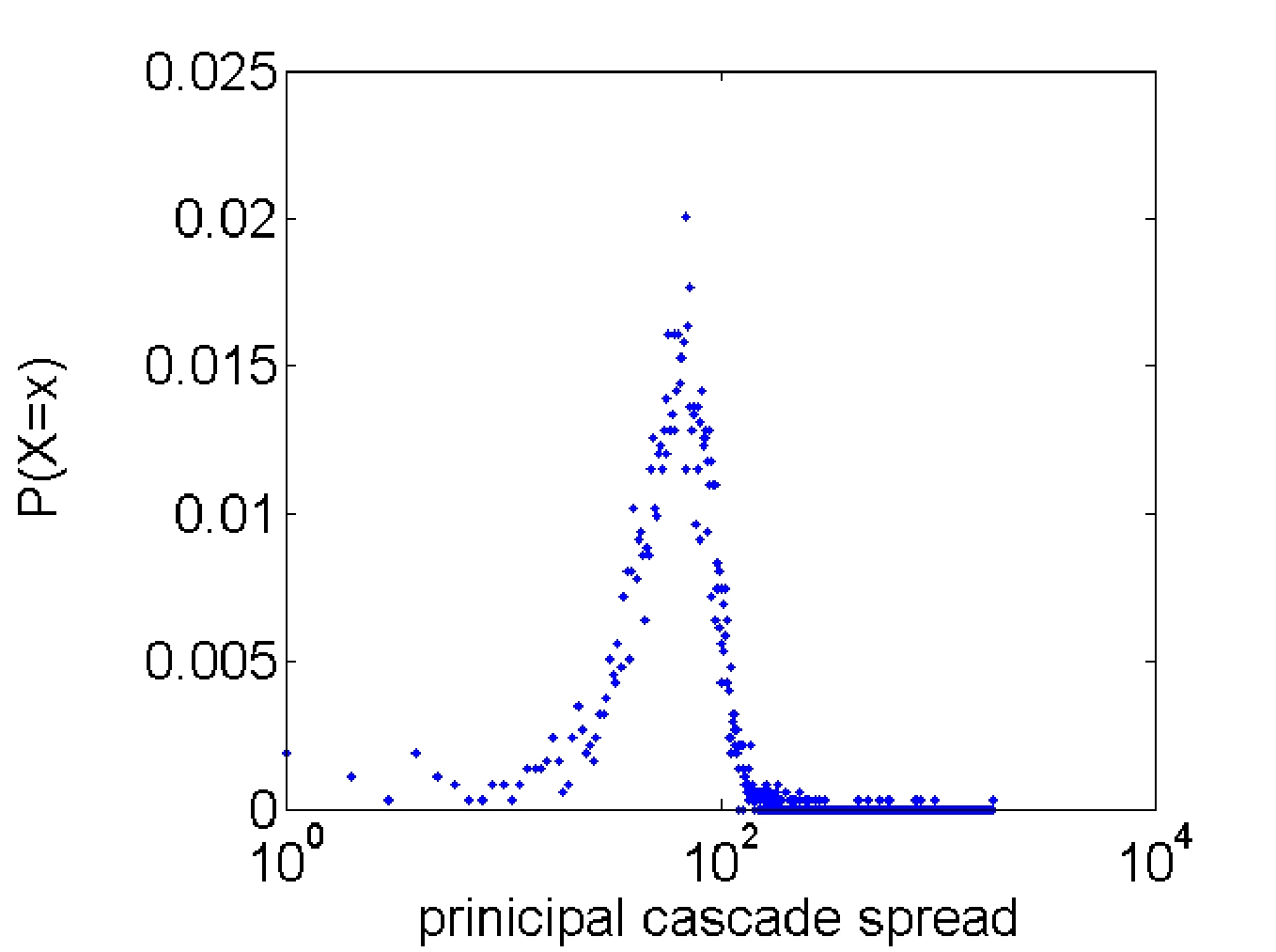} &
  \includegraphics[width=2.5in]{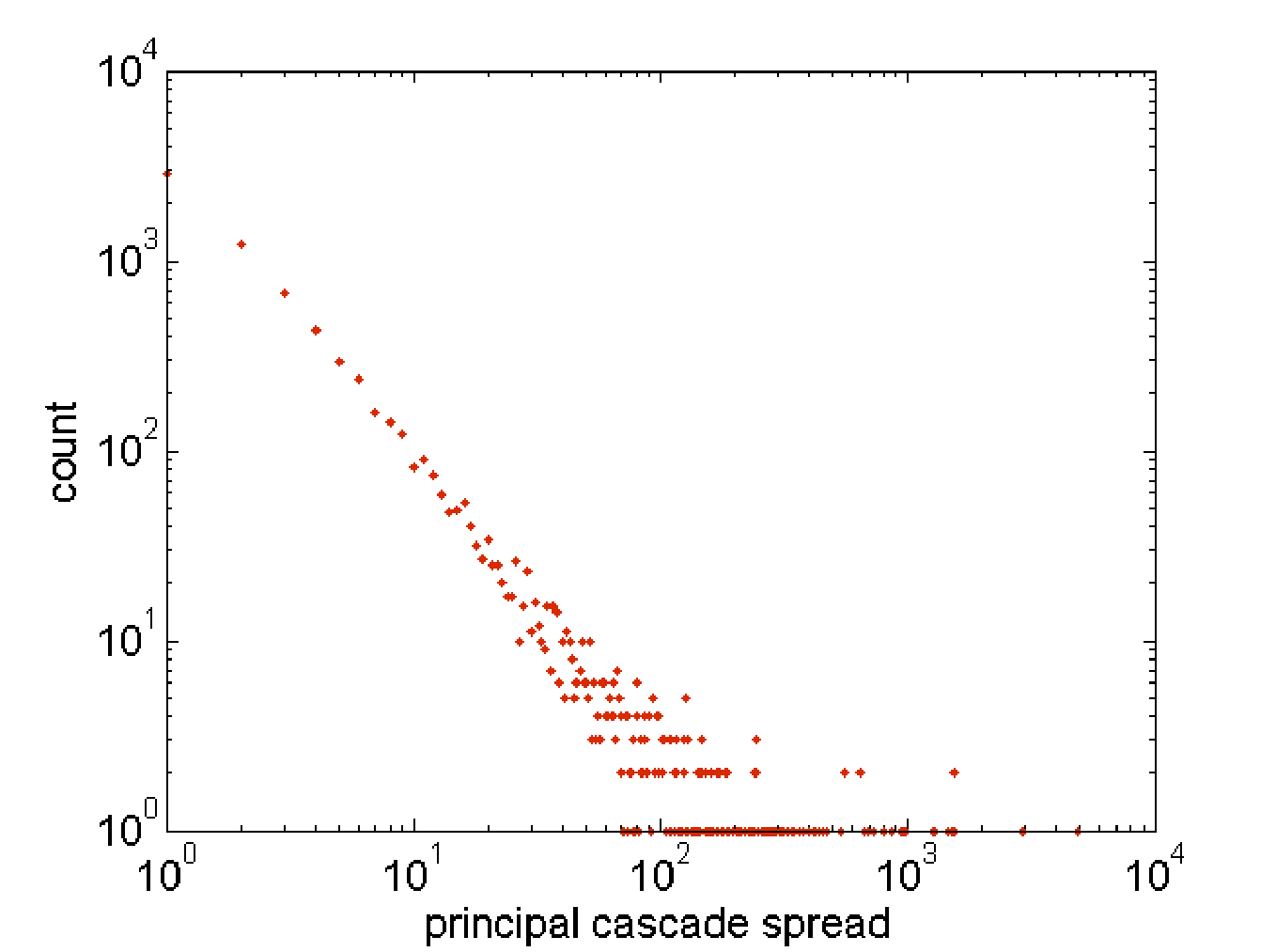} \\
     {(c) Principal cascade spread on Digg } & {(d) Principal cascade spread on Twitter} \\
\end{tabular}
}
\caption{Distribution of spread in Digg and Twitter }
\label{fig:cascade_spread}
\end{figure*}

\subsubsection*{Spread Distribution}
Cascade spread (Fig.~\ref{fig:cascade_spread}) indicates the magnitude of the branching effect. Presence of a fat tail in both Digg and Twitter, both for global spread distribution and principal cascade spread distribution suggests that often a highly connected user, a hub, votes, inducing many followers to vote for the story. Just like cascade size, the peak in Digg's spread distribution likely reflects the influence of top users.

\begin{figure*} [htbp]
\centering{
\begin{tabular}{cc}
  \includegraphics[width=2.5in]{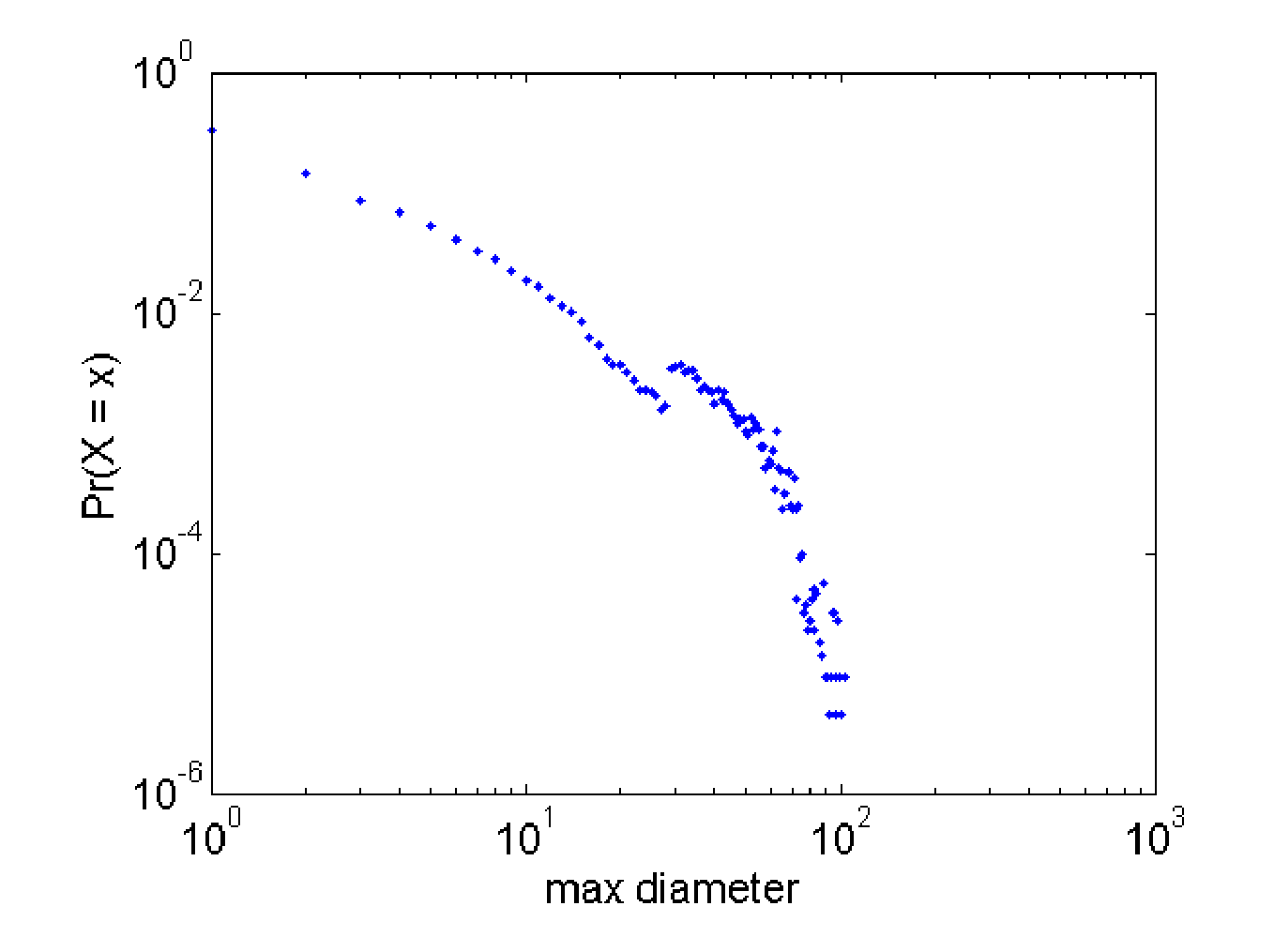} &
 \includegraphics[width=2.5in]{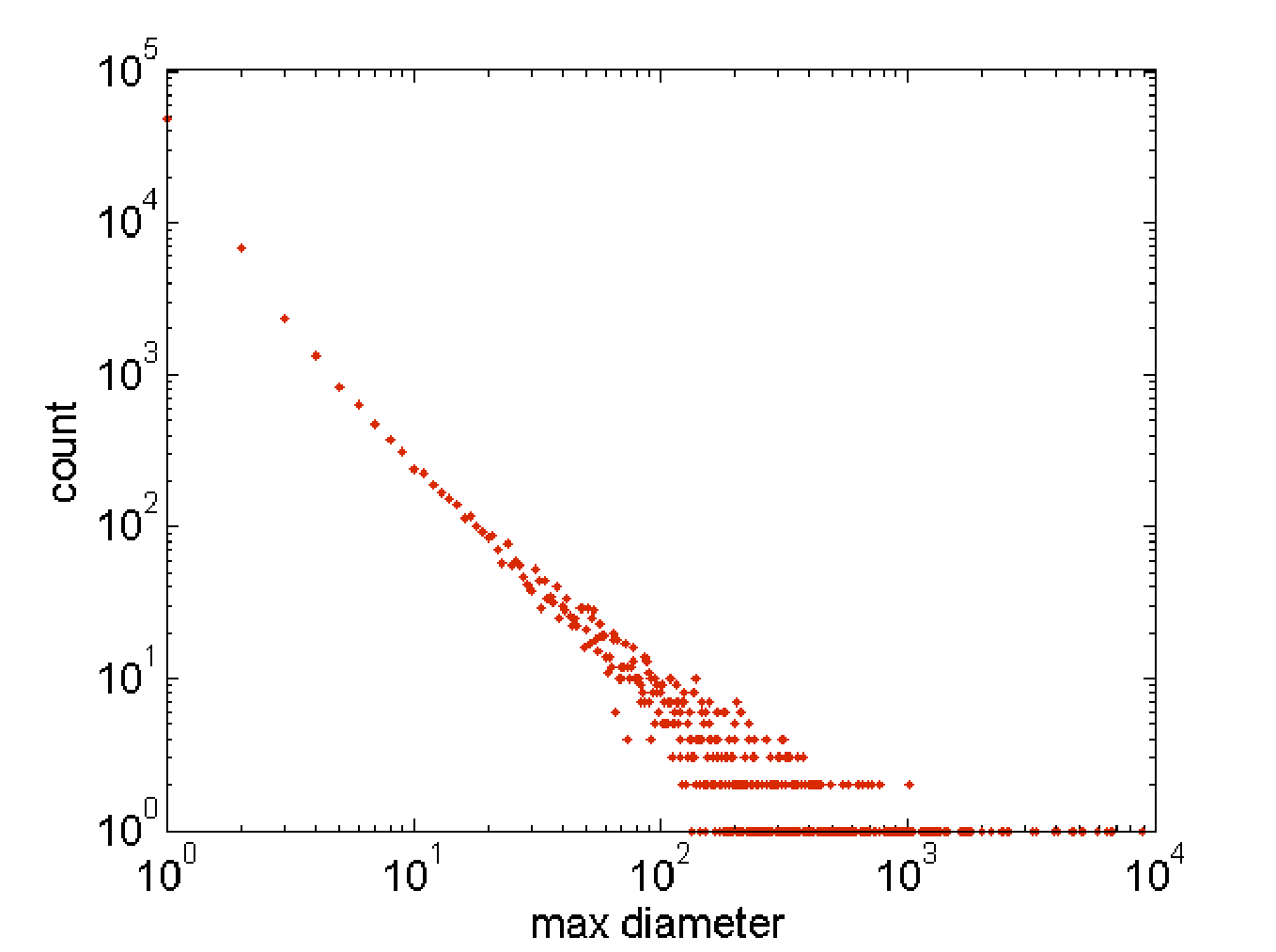} \\
     {(a) Global max. diameter on Digg} &  {(b) Global max. diameter on Twitter}\\
  \includegraphics[width=2.5in]{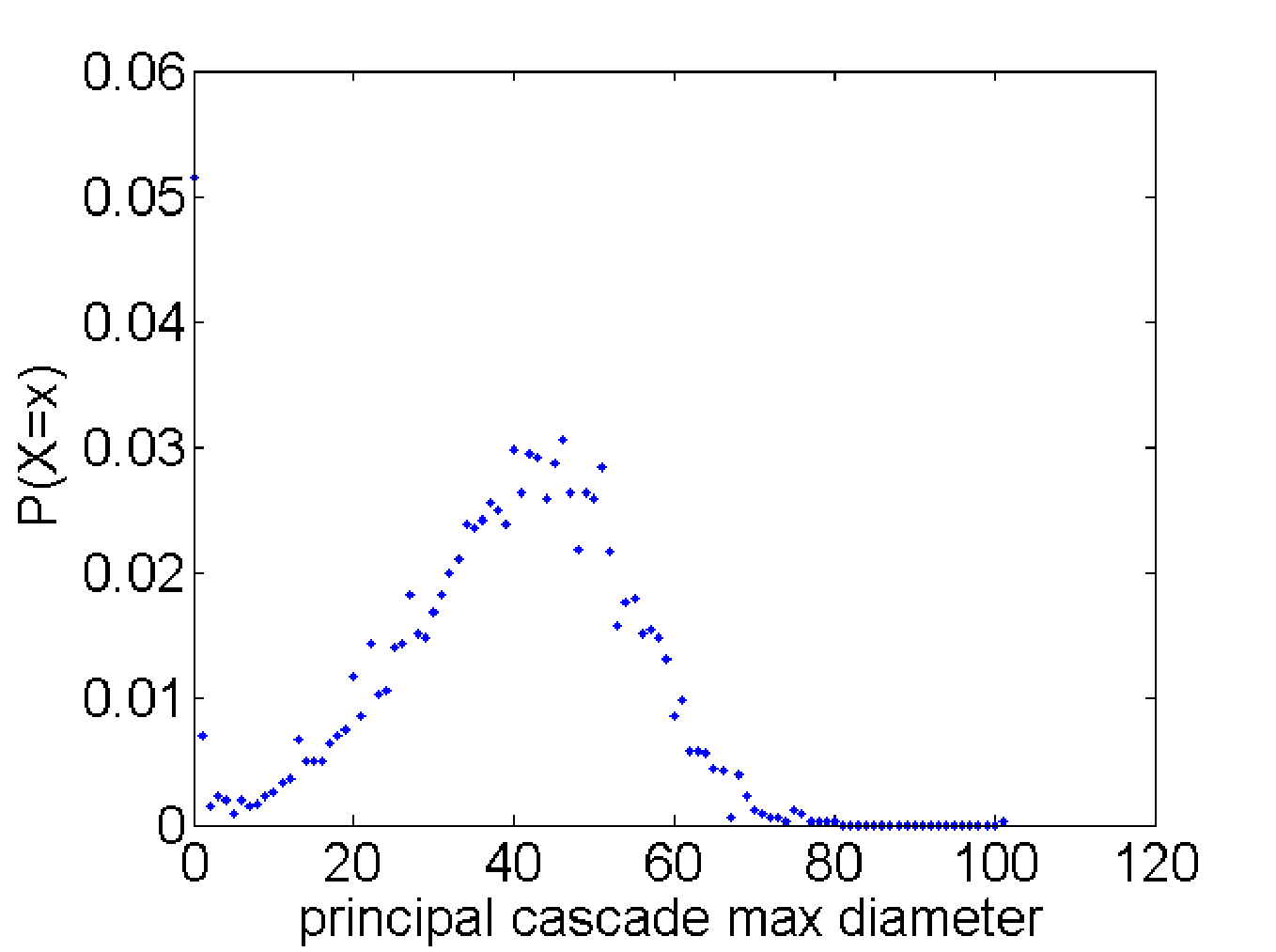} &
  \includegraphics[width=2.5in]{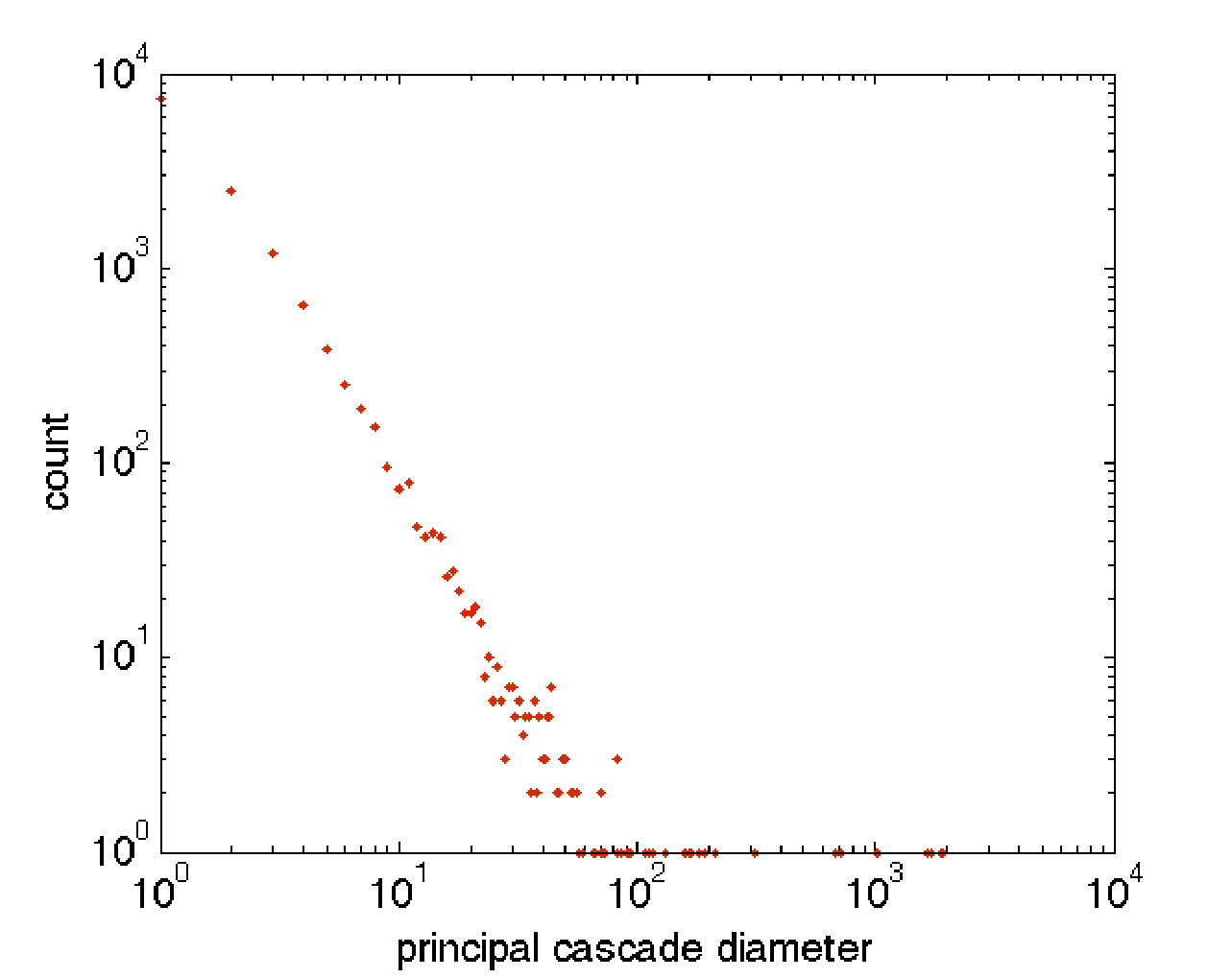} \\
     {(c) Principal cascade max. diameter  on Digg } & {(d) Principal cascade max. diameter on Twitter} \\
\end{tabular}
}
\caption{Distribution of maximum diameter in Digg and Twitter }
\label{fig:cascade_max_diameter}
\end{figure*}

\subsubsection*{Maximum Diameter Distribution}
Similar to cascade size (Fig.~\ref{fig:cascade_size}) and  spread (Fig.~\ref{fig:cascade_spread}) and  maximum diameter (Fig.~\ref{fig:cascade_max_diameter}) has a long tail distribution as on Digg.  However, interestingly,  unlike the rest of  the distributions, the maximum principal cascade diameter in Digg has a normal like distribution, with a mean value of  principal cascade maximum diameter around 40.

\subsubsection*{Minimum Diameter Distribution}
\begin{figure*} [htbp]
\centering{
\begin{tabular}{cc}
  \includegraphics[width=2.5in]{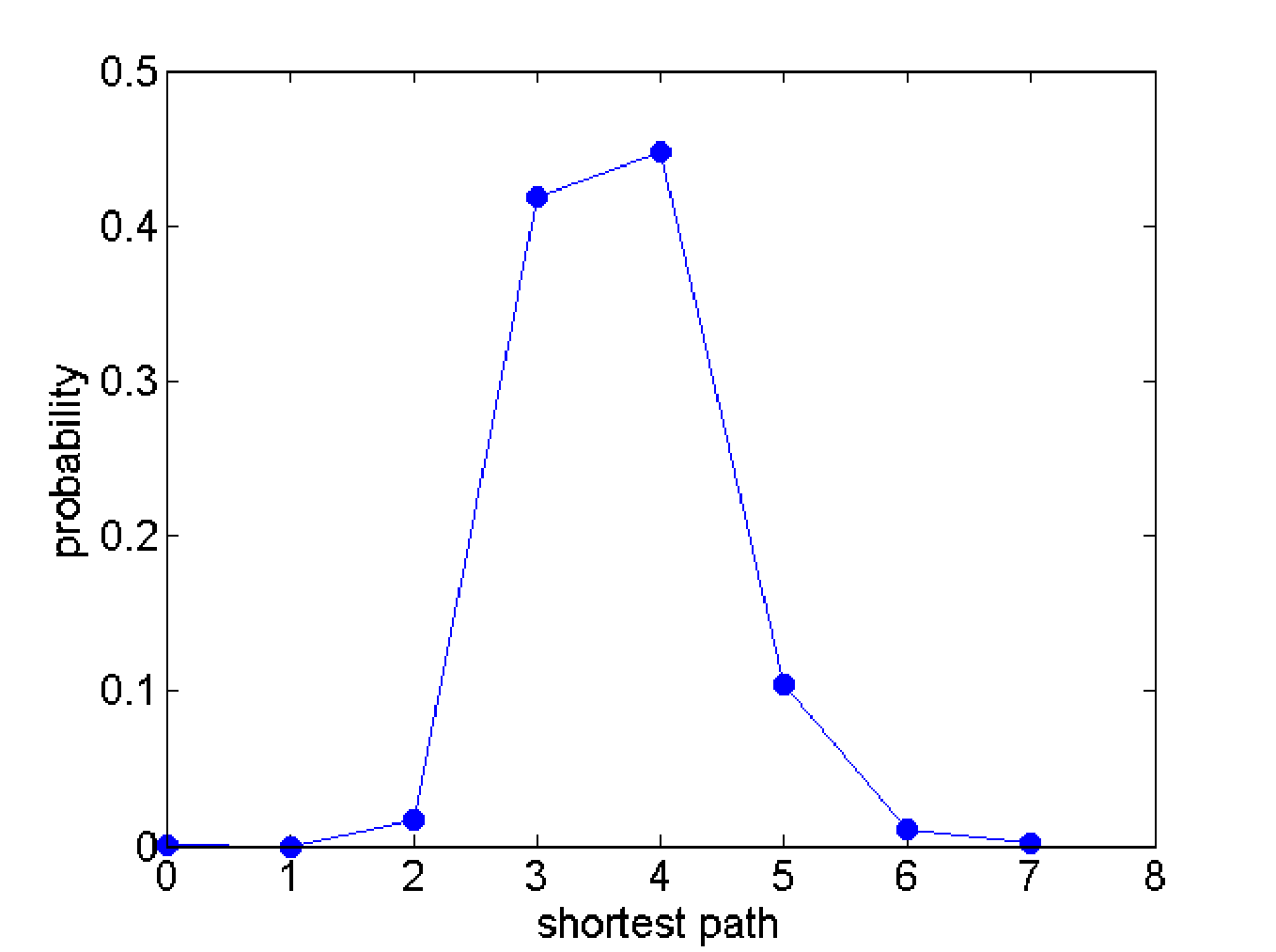} &
 \includegraphics[width=2.5in]{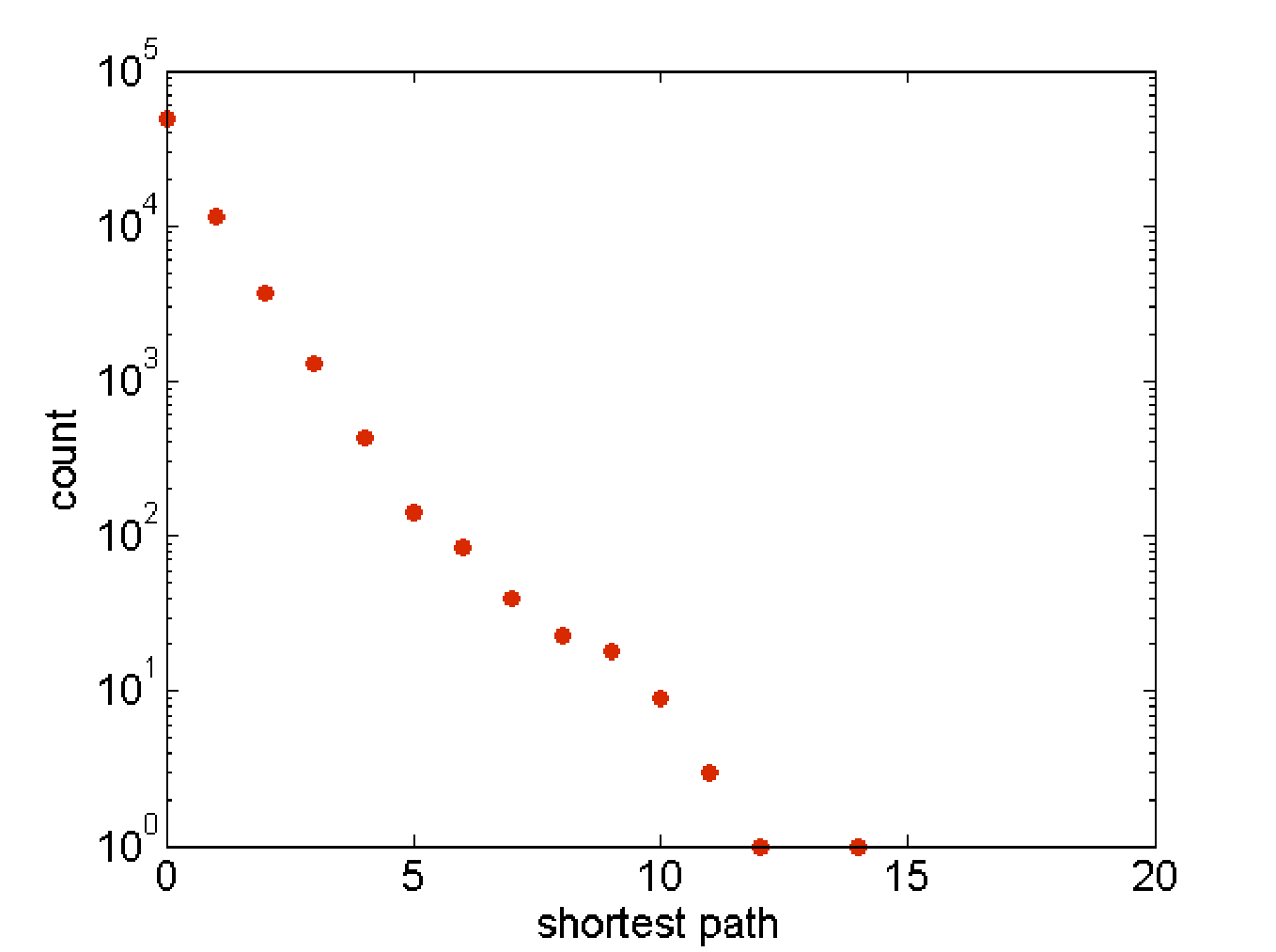} \\
     {(a) Global min. diameter on Digg} &  {(b) Global min. diameter on Twitter}\\
  \includegraphics[width=2.5in]{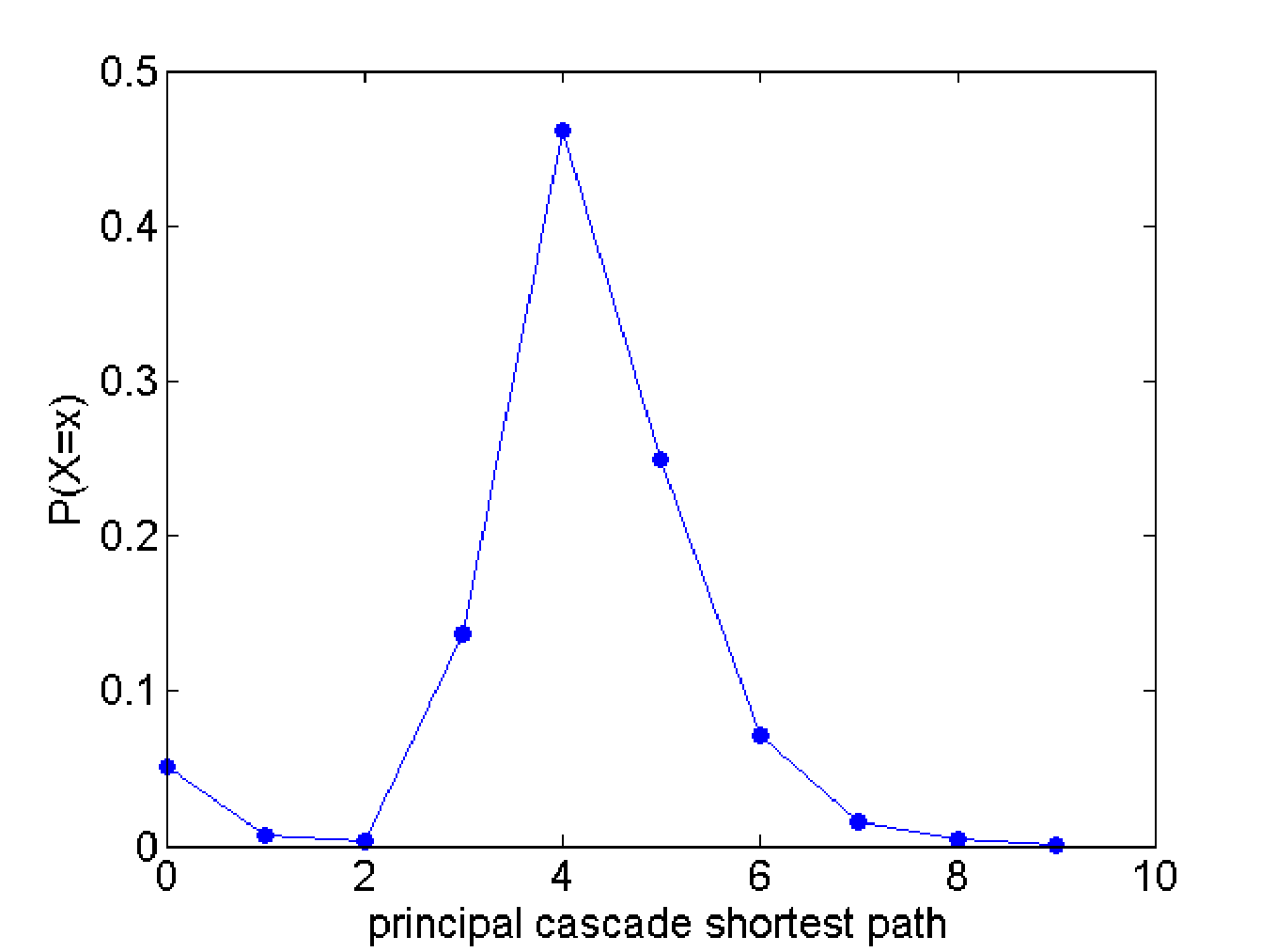} &
  \includegraphics[width=2.5in]{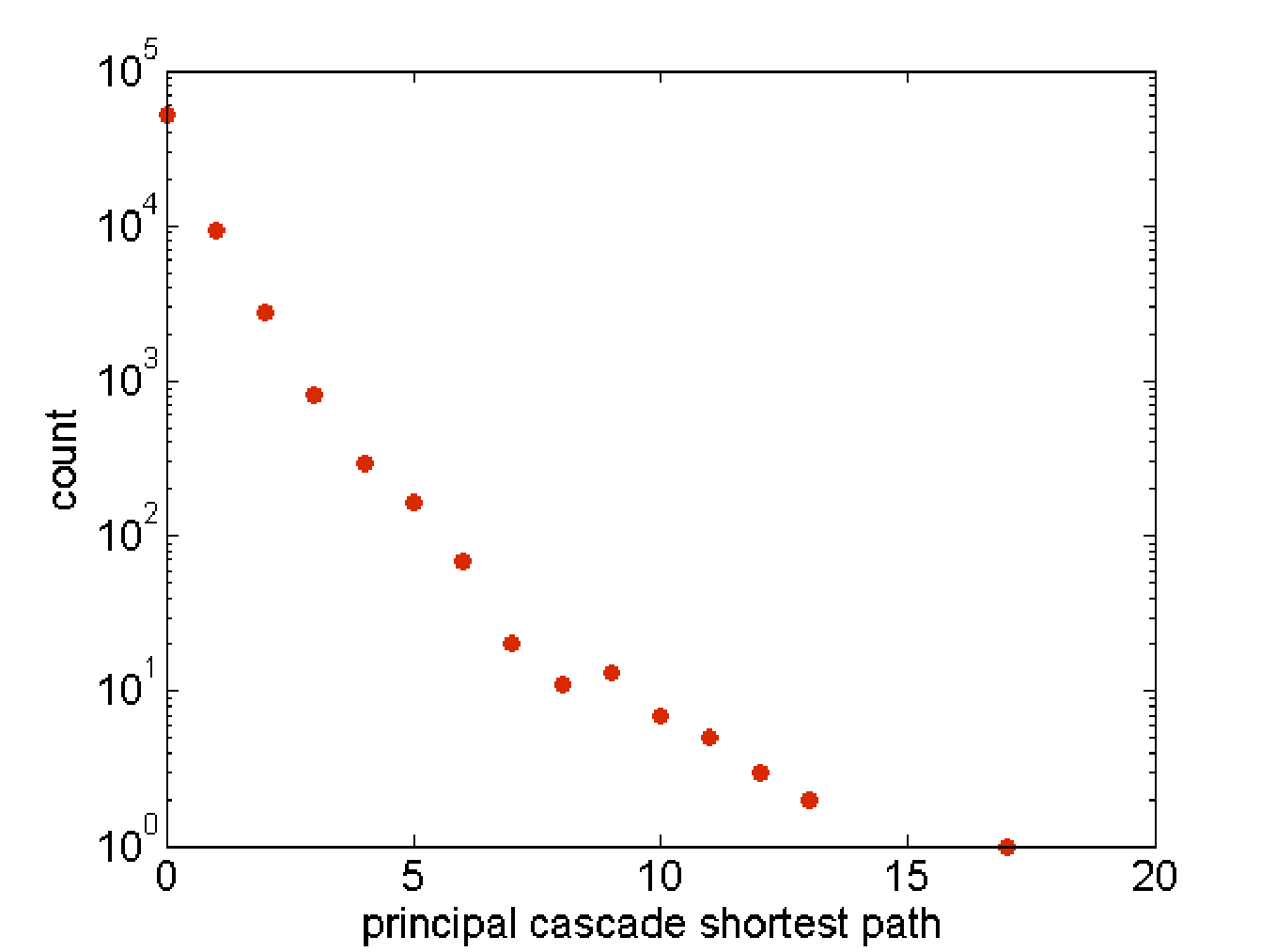} \\
     {(c) Principal cascade min. diameter  on Digg } & {(d) Principal cascade min. diameter on Twitter} \\
\end{tabular}
}
\caption{Distribution of minimum diameter in Digg and Twitter }
\label{fig:cascade_min_diameter}
\end{figure*}
Interestingly, while the global maximum diameter of the cascade on Digg can be quite large (Fig.~\ref{fig:cascade_max_diameter} (a)), the global minimum diameter is at most seven, and often just three or four (Fig.~\ref{fig:cascade_min_diameter}(a)). This could be related to the diameter of the underlying follower graph, although we did not investigate this connection.

However, on Twitter,  the distribution of minimum diameter (Fig.~\ref{fig:cascade_min_diameter}(b) and (d)) looks very different.
The probability of diameter of given length decreases almost monotonically with length. The presence of many small values indicates that many URLs never spread beyond the seed (minimum diameter zero) and its followers (minimum diameter one). A handful of URLs spread more than ten hops from the seed, which though impressively large by the standards of social media, is far shorter than the chains observed in the study of email cascades~\cite{Liben-Nowell08pnas}. One possible explanation is that the email cascades evolved over a longer time period (years), enabling them to grow longer. Although we have observed information cascades on Twitter over a much shorter time period (weeks, rather than years), it is doubtful that they would evolve over a longer time period, given that most of the activity generated by a URL on Twitter takes place within days of submission.

\subsubsection*{Community Effect}
\begin{figure*} [htbp]
\centering{
\begin{tabular}{cc}
  \includegraphics[width=2.5in]{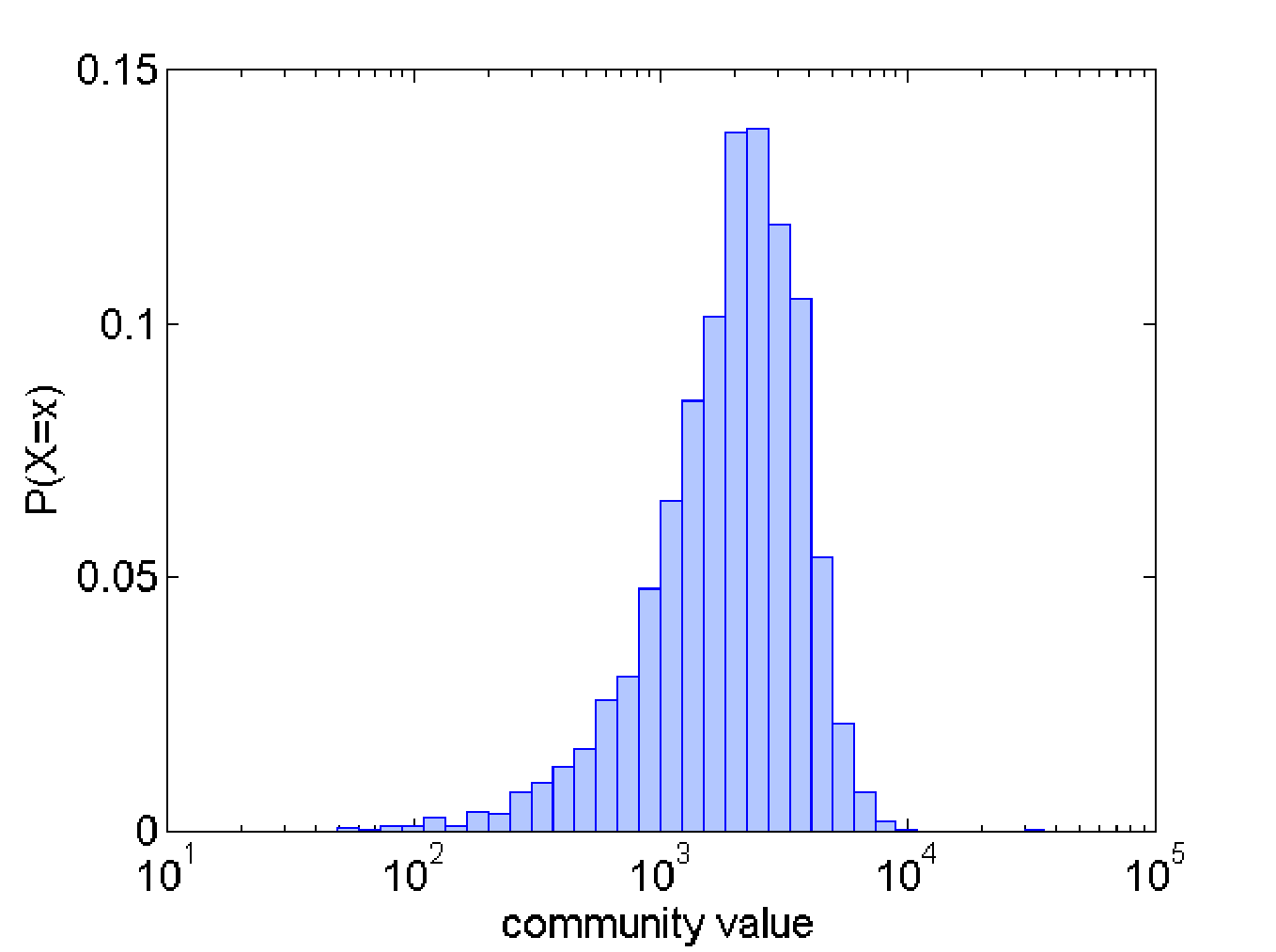} &
 \includegraphics[width=2.5in]{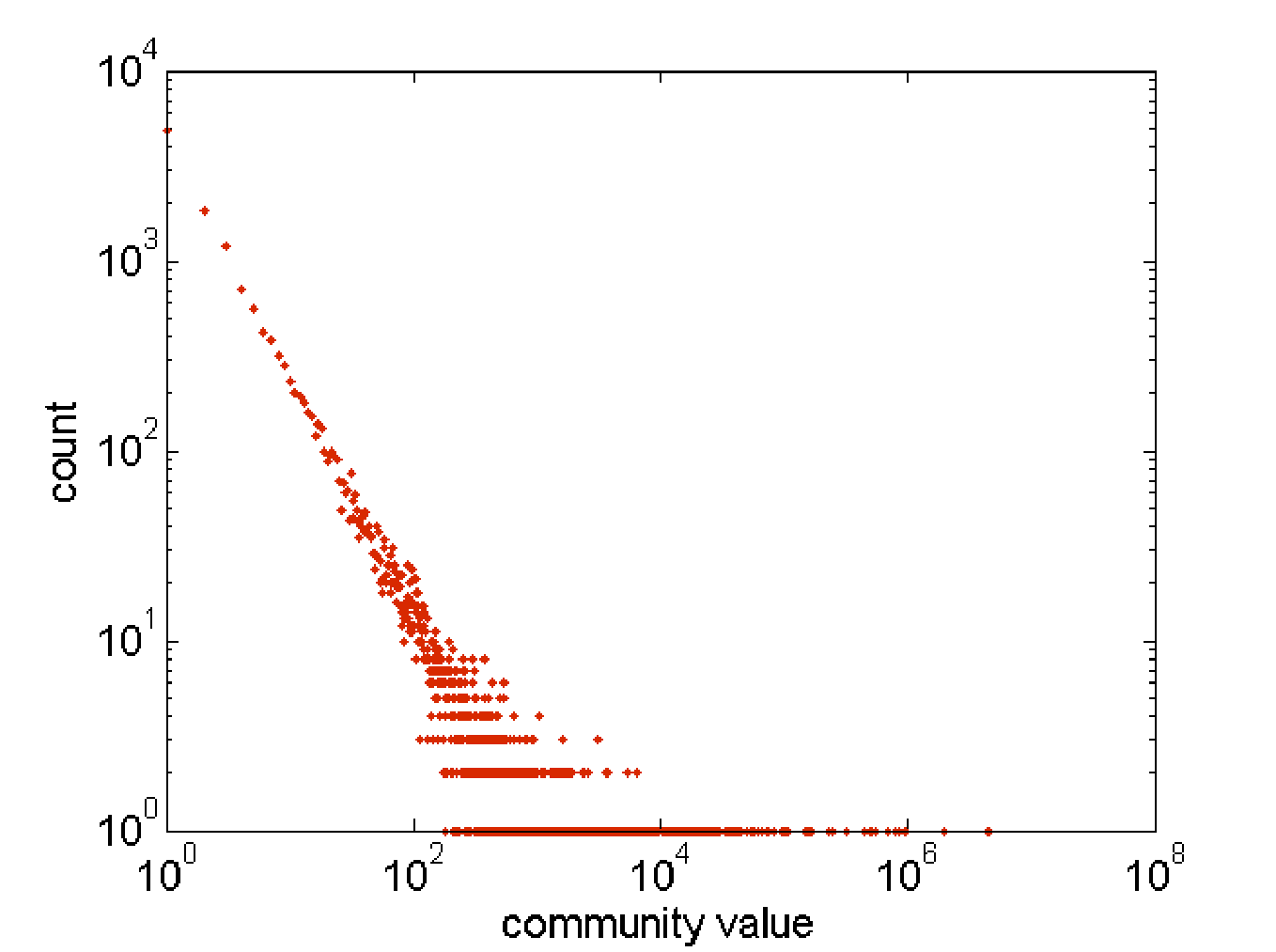} \\
     {(a) Community value on Digg} &  {(b) Community value on Twitter}\\
  \includegraphics[width=2.5in]{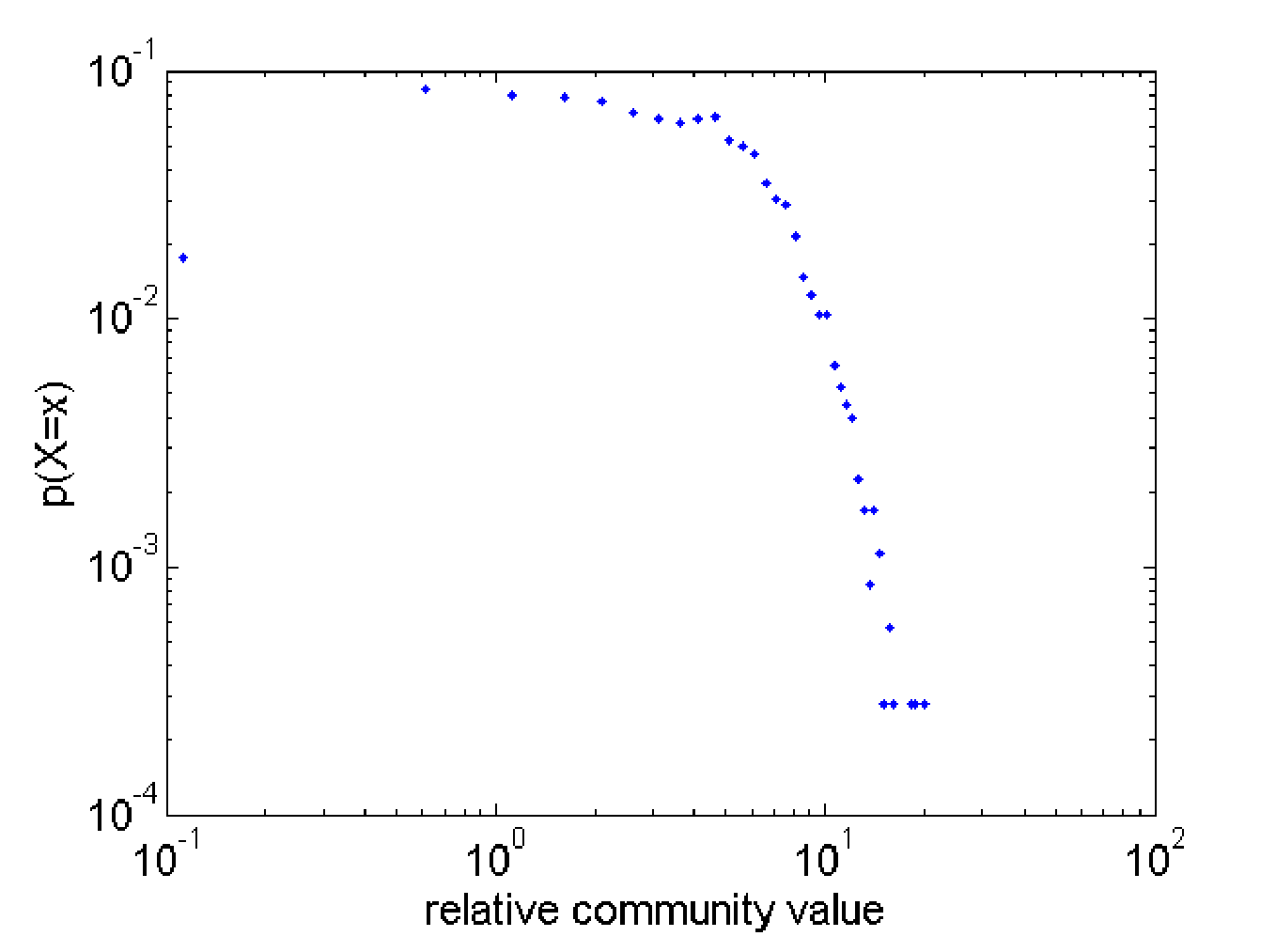} &
  \includegraphics[width=2.5in]{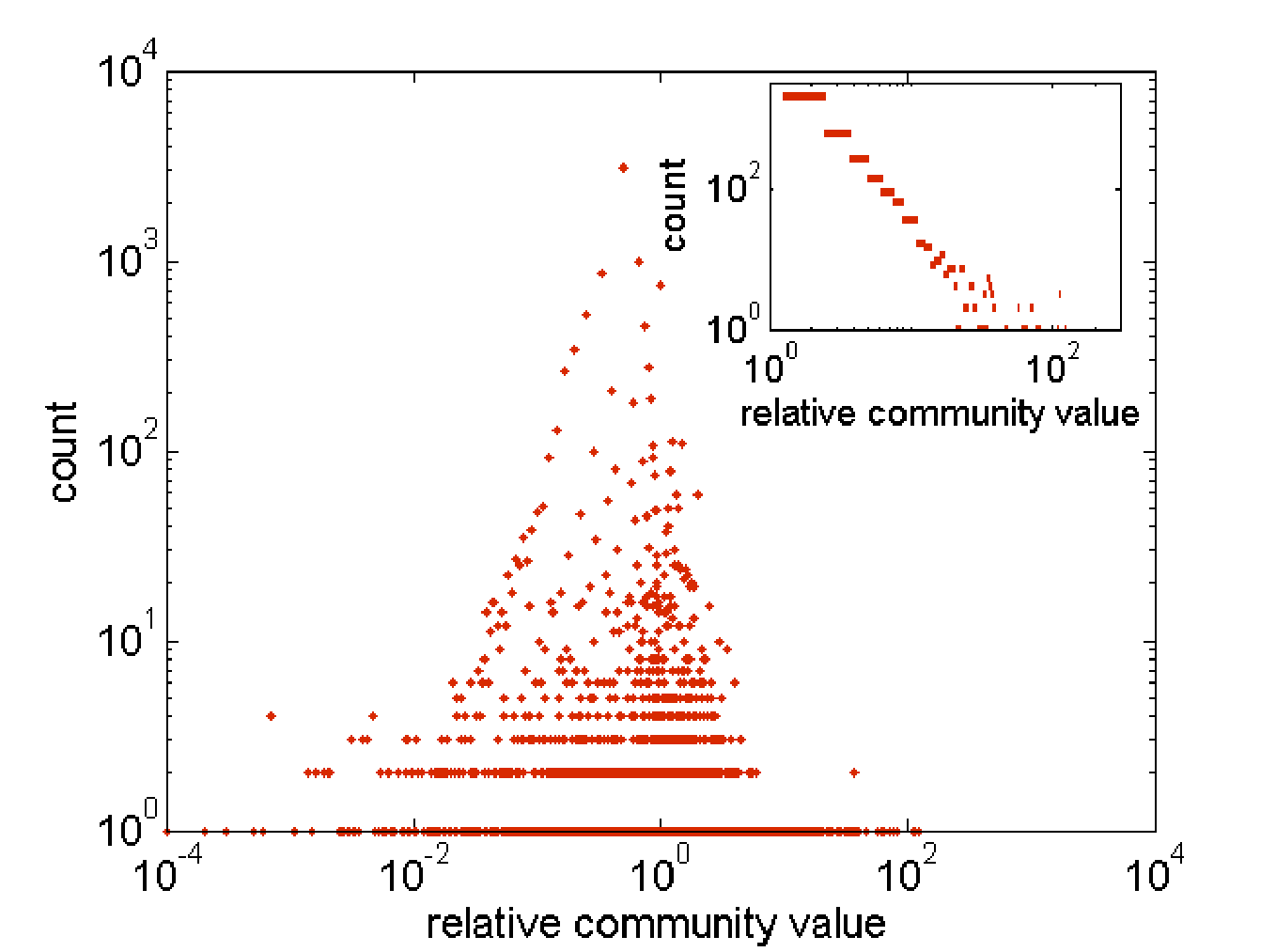} \\
     {(c) Normalized community value on Digg } & {(d) Normalized community value on Twitter} \\
\end{tabular}
}
\caption{Distribution of minimum diameter in Digg and Twitter }
\label{fig:community_val}
\end{figure*}
The community value of Digg cascades (Fig.~\ref{fig:community_val}(a)) displays a lognormal distribution with a maximum around 3,000, suggesting that many cascades spread within a well-connected community. This is further confirmed by normalizing the community value by cascade size (Fig.~\ref{fig:community_val}(c)), which shows that there are many cascades in which each voter follows on average at least ten of the previous voters.

Community value distributions on Twitter (Fig.~\ref{fig:community_val}(b) and (d)) are strikingly different from those on Digg (Fig.~\ref{fig:community_val}(a) \& (c)). Whereas the total community value over all stories on Digg had a lognormal distribution, on Twitter it has a power law-like behavior. We postulate that this difference is due to the structure of follower graphs on Digg and Twitter. Whereas many stories on Digg spread within a community, perhaps even the same community, on Twitter far fewer URLs spread within a community. In fact, each retweeter is  most likely to follow only one previous retweeter, as indicated by the peak at one in the normalized community value distribution (Fig.~\ref{fig:community_val}(d)), suggesting tree-like cascades. A small fraction of URLs do spread within some community, as indicated by large normalized community values in the tail of the distribution. Another interesting observation is about the shape of normalized community value distribution. Whereas the \emph{frequency distribution} has a \emph{log-normal shape}, the \emph{histogram} of normalized community values appears to follow a \emph{power law}\cite{Baek11}.

\subsection*{Discussion}
% Twitter
%17,161,876 mutual links (may have to divide by 2)
%36,743,448 total links
%700K nodes and over 36 million edges.
%0.47

% Digg
%71,834 active users designated
%258,220 friend links.
%125,219 mutual
%279,725 distinct users
%0.48
The cascade properties we measured on Twitter had a scale-free distribution with no characteristic size. These distributions most likely reflect the long tailed distribution of the underlying follower graph. When a highly connected hub joins a cascade, the cascade will branch broadly and increase in size. The hub, however, won't affect the depth of the cascade as much as its spread. Many of the global cascade properties on Digg had a similar scale-free distribution; however, the properties of the principal cascades that started with the submitter had a log-normal distribution. This likely reflects the dominance of top users in the activity of Digg. These users have many followers, especially among other top users, and are responsible for submitting a lion's share of promoted stories~\cite{Lerman07ic,Lerman07sma}. Top users were disproportionately represented in our data set, and the peaks in the distributions of cascade size, etc., are a likely consequence. In other words, when a top user submits a story, he is guaranteed an audience, resulting in a cascade of a certain size. On the other hand, if a poorly connected user submits a story, it will only grow if a well-connected top user picks it up, and few top users follow poorly connected users. Relatively few of the popular stories in our sample were submitted by such users. As we demonstrated in this paper, selection bias that occurs, for example, when Digg promotes stories to the front page, can dramatically affect the shape of the distribution.
Twitter activity, at least as reflected by our data set, is not driven by top users and has less selection bias.

The diameter and community value distributions suggest a difference in the structure of the follower graphs on which cascades are spreading. As the previous study suggested~\cite{Lerman10icwsm}, Digg follower graph is dense and tightly interconnected, with an underlying community structure, at least among top users. Twitter graph, on the other hand, does not appear to have significant community structure. While many Digg cascades spread within such tightly connected communities, with each node (voter) connected on average to several previous voters, Twitter cascades appear to be more ``dendritic'' or tree-like, with each node following on average one previous tweeter. Community structure could also explain the difference in principal cascade size distribution. Many of the stories in our sample were submitted by top users, who form a community with other top users. The size of the cascade most likely is explained by the size of the community.

Despite differences in the size and structure of the underlying follower graph, and how the content is featured on these sites, Digg and Twitter cascades look remarkably similar. On both networks, though information cascades spread fast enough for one seed to infect thousands of users, they end up affecting less than 1\% of the follower graph. This is in contrast to our understanding of the dynamics of epidemics on graphs \cite{Wang03}, which suggests the existence of an epidemic threshold  above which epidemics spread to a significant fraction of the graph.
In recent study of Digg~\cite{Versteeg11icwsm} we demonstrated that  two complementary effects limit the final size of cascades.
First, because of the highly clustered structure of the Digg network, most people who are aware of a story have been exposed to it via multiple friends.  This functions to lower the epidemic threshold while also slowing the growth of cascades.
We also found that the social contagion mechanism on Digg deviates from standard social contagion models, like the independent cascade model, and this severely curtails the size of social epidemics on Digg. In fact, these findings underscore the fundamental difference between information spread and other contagion processes: despite multiple opportunities for infection within a social group, people are less likely to become spreaders of information with repeated exposure.
It is an open question whether the same mechanism applies to Twitter cascades.

Our work suggests a possible explanation to the deep and narrow chains in email forwarding cascades observed by \cite{Liben-Nowell08pnas}. This study reconstructed cascades from the signatures on the forwarded email petitions.
%Thus the observed cascade would have a tree like structure.
This method offers only a partial view of the network and does not identify all edges between individuals that participated in the email chain, because an individual could have received multiple emails, but will respond only to one.  If an individual has already forwarded the message, she will not do so again, and an edge between her and the sender will not be observed.  As shown in Figures~\ref{fig:cascade_max_diameter}--\ref{fig:cascade_min_diameter}, though the minimum diameter is relatively small, the maximum diameter of some cascades is quite large. If we represent each cascade as a graph and sample a tree, by randomly picking one of the activation edges each node (if it has several activation edges), the resulting tree is likely to be deep and narrow.
%However, if some of the activation edges in the cascade are not observed, as would be the case when the network is reconstructed from partial information present in the cascades (as in the email chain letter study), the diameter observed might be greater than the actual minimum diameter and the observed spread might be narrower than the actual spread.
Therefore, missing information may lead to a different \emph{observed} cascade structure compared to the \emph{actual} structure.

% Conclusions
% are cascades treelike? are they community like?
% Do we need to exclude non-news stories from Twitter?

\section*{Related Work}
\label{sec:related}
Previous empirical studies of information cascades produced conflicting results. \cite{Wu03} examined patterns of email forwarding within an organization and found that email forwarding chains terminate after an unexpectedly small number of steps. They argued that unlike the spread of a virus on a social network, the flow of information is slowed by decay of similarity among individuals within the social network. They measured similarity by distance within the organizational hierarchy between the two individuals. Similarly, a large-scale study of the effectiveness of word-of-mouth product recommendation~\cite{Leskovec06} found that most recommendation chains terminate after one or two steps.
%However, authors noted sensitivity of recommendation to price and category of product,  leaving open the question whether social networks are an effective tool for disseminating information, rather than purchasing products.
\cite{Leskovec07blogs} studied the structure of cascades formed by hyperlinks between blog posts. \cite{Moon10} used a similar methodology to study of information cascades on Twitter. Both studies enumerated common cascade shapes, including star and chain, and provided their occurrence statistics. They found chains to be at most of length ten, with the spread having a long tail distribution ranging up to hundreds of nodes.
Contrary to these findings, \cite{Liben-Nowell08pnas} found that email forwarding cascades produced by two popular email petitions were extremely deep (long chains) and narrow (low spread).

%Contrary to these studies, we find that information, such as news, reaches many individuals within a social network. Moreover, the reach of information spread does not seem to depend on similarity between users, at least when similarity is measured by number of edges between them. On Digg, whose users are highly interconnected, a  story does not reach as many fans as on Twitter, where users are less densely connected.

In all of these studies, however, the structure of the underlying network was not directly visible but had to be inferred by observing linking or email forwarding behavior. In our study, on the other hand, the networks are extracted independently of data about the spread of information. This helps us to get a more accurate representation of how information spreads in online social networks, since we are able to take into account the edges and nodes that would be otherwise missed, when the network is inferred from the information spread as discussed in the previous section.

% blog space
%A number of researchers have studied the flow of information and influence in the blogosphere and in a virtual world. \cite{Gruhl04} traced topic propagation through blogs and used a model of the spread of epidemics on networks~\cite{newman02} to characterize the spread of topics through the blogosphere.
%\cite{Leskovec07blogs} defined an information cascade as a graph of hyperlinks between blog posts. A cascade starts with a cascade initiator, with other blog posts joining the cascade by linking to the initiator or other members of the cascade. Leskovec et al. found that the distribution of cascade sizes follows a power law. In these studies, the networks were derived from the observed links between blog posts, i.e., from the diffusion of information. In our study, on the contrary, they were extracted from the sites independently of data about the diffusion of information. \cite{bakshy09} traced the spread of influence in a multi-player online game and found that similar to our findings with social news, influence spreads easily on social networks in virtual worlds. This provides an independent confirmation of the importance of social networks in the dynamics of information flow.

In a previous work, we proposed a methodology to quantitatively characterize the microscopic and macroscopic structure of information cascades and used it to study evolution of cascades on Digg~\cite{Ghosh11wsdm}. In this study, we use this methodology for comparative analysis of the macroscopic properties of cascades on Digg and Twitter. We also introduce a new macroscopic features which quantifies the effect of community, and show that community structure of the network affect information spread.

\section*{Conclusion}
\label{sec:conclusion}
We conducted an empirical analysis of user activity on Digg and Twitter. Though the two sites are have different functionality and user interface, they are used in strikingly similar ways to spread information. On both sites users actively create social networks by creating links to people whose activities they want to follow. Users employ these networks to discover interesting information that they then spread to other by voting for it on Digg or retweeting it on Twitter. In spite of the similarities, there are quantitative differences in the user interface and the structure networks on Digg and Twitter, and these differences affect how far and how quickly information spreads. Digg networks are dense and highly interconnected~\cite{Lerman10icwsm} and many of the cascades appear to spread through an interconnected community. Twitter cascades, on the other hand, are more tree-like.
%A newly-posted story on Digg spreads mainly through the network, as submitter's followers and their own followers vote on the story. After the story is promoted to Digg's front page, however, it is visible to a large number of unconnected users and continues to spread from new seeds.
%The spread of the story on the network slows significantly, though the story may still generate a large response from Digg audience.
%Twitter does not have a similar promotion mechanism, and information there spreads mainly through the network. Although Twitter's network is less structured than Digg's, information penetrates its network more than Digg's.
%On both sites, cascades are shallow but not as deep and narrow as observed in other studies of information spread~\cite{Liben-Nowell08pnas}.

% model of social dynamics, quality
Understanding characteristics of user activity and the effect networks have on it is especially critical for the effective use of social media and peer production systems. Currently these systems aggregate over activities of many people to identify trending topics and noteworthy contributions.  Most of these sites also highlight activities of others within a person's social network. Since people create social links to others who are similar to them, or whose contributions they find interesting, the dynamics of information spread in a network may be different from its spread outside the network. Separating in-network from out-of-network activity allows us, among other things, to better estimate the inherent quality of the contributions~\cite{CraneSornette08} or predict their future activity~\cite{Lerman08wosn,Hogg10icwsm,Lerman10www}.

% Use the PLoS provided bibtex style
%\bibliographystyle{plos}
%\bibliography{references}

\end{document}